\documentclass[utf8]{aa}  
\usepackage{graphicx}
\usepackage{txfonts}
\usepackage[utf8]{inputenc}
\usepackage{natbib}
\usepackage{caption}
\usepackage{subcaption}
\usepackage{amsmath}
\usepackage{siunitx}
\usepackage[normalem]{ulem}
\usepackage[usenames,dvipsnames]{color}
\usepackage[usenames,dvipsnames,svgnames]{xcolor}
\usepackage{array}
\usepackage{bm}
\usepackage[hidelinks]{hyperref}
\defcitealias{2017ApJ...841...30T}{PT17}
\defcitealias{10.1111/j.1365-2966.2011.18315.x}{H11}
\bibpunct{(}{)}{;}{a}{}{,}
\addto\extrasenglish{
}

\newcommand{\ed}{}

\begin{document}

   \title{Idealised simulations of the deep atmosphere of hot jupiters:}

%   \title{Exploring the Internal Structures of hot Jupiters using the GCM DYNAMICO}
   \subtitle{deep, hot, adiabats as a robust solution to the radius inflation problem }

   \author{F. Sainsbury-Martinez\inst{1}\thanks{\email{felix.sainsbury@cea.fr}} \and P. Wang\inst{2,1} \and S. Fromang\inst{4} \and P. Tremblin\inst{1} \and T. Dubos\inst{5} \and Y. Meurdesoif\inst{6} \and A. Spiga\inst{5,7} \and J. Leconte\inst{7} \and I. Baraffe\inst{8,3} \and G. Chabrier\inst{3,8} \and N. Mayne\inst{8} \and B. Drummond\inst{8} \and F. Debras\inst{9}  \fnmsep}

   \institute{Maison de la Simulation, CEA, CNRS, Univ. Paris-Sud, UVSQ, Université Paris-Saclay\ \and Univ. Lyon, ENS de Lyon, Univ. Claude Bernard, CNRS, Laboratoire de Physique, Lyon, France\ \and Ecole Normale Superieure de Lyon, CRAL, UMR CNRS 5574 \ \and Laboratoire AIM, CEA/DSM-CNRS-Université Paris 7, Irfu/Departement d'Astrophysique, CEA-Saclay, 91191 Gif-sur-Yvette, France \ \and Laboratoire de Météorologie Dynamique (LMD/IPSL), Sorbonne Université, Centre National de la Recherche Scientifique, École Polytechnique, École Normale Supérieure, Paris \ \and Laboratoire des Sciences du Climat et de l'Environnement/Institut Pierre-Simon Laplace, Université Paris-Saclay, CEA Paris-Saclay \ \and Laboratoire d'astrophysique de Bordeaux, Univ. Bordeaux, CNRS, B18N, allée Geoffroy Saint-Hilaire, 33615 Pessac, France. \and Astrophysics Group, University of Exeter, Exeter, Devon \and IRAP, Université de Toulouse, CNRS, UPS, Toulouse, France \
}

   \date{Received 02 August 2019; accepted 14 Nov 2019}

% \abstract{}{}{}{}{} 
% 5 {} token are mandatory
  \abstract
   {The anomalously large radii of hot Jupiters has long been a mystery. However, by combining both theoretical arguments and 2D models, a recent study has suggested that the vertical advection of potential temperature leads to an adiabatic temperature profile in the deep atmosphere hotter than the profile obtained with standard 1D models.}
   {In order to confirm the viability of that scenario, we extend this investigation to three dimensional, time-dependent, models.}
 {We use a 3D General Circulation Model (GCM), DYNAMICO to perform a series of calculations designed to explore the formation and structure of the driving atmospheric circulations, and detail how it responds to changes in both the upper and deep atmospheric forcing.} 
   {In agreement {\ed with the previous, 2D, study}, we find that a hot adiabat is the natural outcome of the long-term evolution of the deep atmosphere. Integration times of order $1500$ years are needed for that adiabat to emerge from an isothermal atmosphere, explaining why it has not been found in previous hot Jupiter studies. Models initialised from a hotter deep atmosphere tend to evolve faster toward the same final state. We also find that the deep adiabat is stable against low-levels of deep heating and cooling, as long as the Newtonian cooling time-scale is longer than $\sim 3000$ years at $200$ bar.}
   {We conclude that the steady-state vertical advection of potential temperature by deep atmospheric 
     circulations constitutes a robust mechanism to explain hot Jupiter inflated
     radii. We suggest that future studies of hot Jupiters are evolved for a longer time than currently done, and, when possible, include models initialised with a hot deep adiabat. 
    We stress that this mechanism stems from the advection of entropy by irradiation induced mass flows and does not require (finely tuned) dissipative process, in contrast with most previously suggested scenarios.}


   \maketitle
%
%-------------------------------------------------------------------

% In general, captions are too long. They should only describe the meaning the plotted symbols. The physical description & interpretation should be in the text.
\section{Introduction} 
\label{sec:introduction}
The anomalously large radii of highly irradiated Jupiter-like exoplanets, known as hot Jupiters, remains one of the key unresolved issues in our understanding of extrasolar planetary atmospheres. The observed correlation between the stellar irradiation of a hot Jupiter and its observed inflation \citep[for examples, see][]{Demory_2011,2011ApJ...729L...7L,2016ApJ...818....4L,2018A&A...616A..76S} suggests that it is linked to the amount of energy deposited in the upper atmosphere. Several mechanisms have been suggested as possible  explanations (see \citealt{Baraffe_2009,2014prpl.conf..763B,2010SSRv..152..423F}, for a review). These solutions include tidal heating and physical (i.e. not for stabilisation reasons) dissipation (\citealt{refId0999,2010ApJ...714....1A,2019MNRAS.484.5845L}), ohmic dissipation of electrical energy (\citealt{Batygin_2010,2010ApJ...719.1421P,2011ApJ...738....1B,2012ApJ...750...96R}), deep deposition of kinetic energy (\citealt{2002A&A...385..156G}), enhanced opacities which inhibit cooling \citep{Burrows_2007} or ongoing layered convection that reduces the efficiency of heat transport \citep{Chabrier_2007}. At present time, however, there is no consensus across the community on a given scenario because {\ed the majority} of these solutions require finely tuned physical environments which either deposit additional energy deep within the atmosphere or affect the efficiency of vertical heat transport. \\
Recently, \citet{2017ApJ...841...30T}, hereafter \citetalias{2017ApJ...841...30T}, suggested a mechanism that naturally arises from first physical principles. Their argument goes as follows: consider the equation for the evolution of the potential temperature $\Theta$, which is equivalent to entropy in this case:
\begin{equation}
\frac{D\Theta}{Dt} = \frac{\Theta H}{Tc_{p}} \, , 
\label{eq:2ndlaw}
\end{equation}
where $D/Dt$ stands for the Lagrangian derivative in spherical coordinates, $H$ is the local heating or cooling rate, $c_{p}$ is the heat capacity at constant pressure, and $\Theta$ is defined as a function of the temperature $T$ and pressure, $P$:
\begin{equation}
\Theta = T\left(\frac{P_0}{P}\right)^{\frac{\gamma-1}{\gamma}} \, ,
\end{equation}
where $P_{0}$ is a reference pressure and $\gamma$=$C_p/C_v$ is the adiabatic index. In a steady state, \autoref{eq:2ndlaw} reduces to 
\begin{equation}
\bm{u}\cdot\nabla\Theta = \frac{\Theta H}{Tc_{p}} \, , 
\label{eq:Y}
\end{equation}
where $\bm{u}$ is the velocity. In the deep atmosphere, radiative heating and cooling both tend to zero (i.e. $H\rightarrow 0$) because of large atmospheric opacities. {\ed In this case (with $H\rightarrow 0$), and} if the winds do not vanish (i.e. $\left|\bm{u}\right|\neq 0$, see \autoref{sec:main_results}), the potential temperature $\Theta$ must remain constant for \autoref{eq:Y} to be valid. In other words, the temperature-pressure profile must be adiabatic and satisfy the scaling:
\begin{equation}
P \propto T^{\frac{\gamma}{\gamma-1}} \, .
\end{equation} 
We emphasise that this adiabatic solution is an equilibrium  that does not require any physical dissipation. {\ed There is an internal energy transfer to the deep atmosphere, through an enthalpy flux, but there is no dissipation from kinetic, magnetic, or radiative energy reservoirs to the internal energy reservoir.} Dissipative processes $D_\mathrm{dis}$ would act as a source term with $\bm{u}\cdot \nabla \Theta \propto D_\mathrm{dis}$ and would drive the profile away from the adiabat.  \\
Physically, as discussed by \citetalias{2017ApJ...841...30T}, this constant potential temperature profile in the deep atmosphere  is driven by the vertical advection of potential temperature from the outer and highly irradiated atmosphere to the deep atmosphere {\ed by large scale dynamical motions} where it is {\ed almost completely} homogenised by the residual global circulations (which themselves can be linked to the conservation of mass and momentum, and the large mass/momentum flux the super-rotating jet drives in the outer atmosphere). The key point is that it causes the temperature-pressure profile to converge to an adiabat at lower pressures than those at which the atmosphere becomes unstable to convection. As a result, the outer atmosphere connects to a hotter internal adiabat than would be obtained through a standard, 'radiative-convective' single column model. This potentially leads to a larger radius compared with the predictions born out of these 1D models.\\
Whilst \citetalias{2017ApJ...841...30T} was able to confirm this hypothesis
through the use of a 2D stationary circulation model, there are still a number
of limitations to their work. Maybe most importantly, the models they used only
considered the formation of the deep adiabat within a 2D equatorial slice. The
steady-state temperature-pressure profiles at other latitudes remains unknown,
as well as the nature of the global circulations at these high pressures in the
equilibrated state. Strong ansatzes were also made about the nature of the
meridional (i.e. vertical and latitudinal) wind at the equator, with their models prescribing the ratio of latitudinal to vertical mass fluxes, that could potentially affect the proposed scenario. The purpose of this paper is to reduce and constrain these assumptions and limitations and to demonstrate the viability of a deep adiabat at equilibrium. This is done by means of a series of idealised 3D GCM calculations designed such as to allow us to fully explore the structure of the deep atmospheric circulations in equilibrated hot Jupiter atmospheres, as well as investigate the time-evolution of the deep adiabat. As we demonstrate in this work, the adiabatic profile predicted by \citetalias{2017ApJ...841...30T} naturally emerges from such calculations and appears to be robust against changes in the deep atmosphere radiative properties. This is the core result of this work.\\
The structure of the work is as follows. Our simulations properties are
described in \autoref{sec:method}, where we introduce the GCM DYNAMICO, used
throughout this study. We then demonstrate that, when using DYNAMICO, not only
are we are able to recover standard features observed in previous
short-timescale studies of hot Jupiter atmospheres
(\autoref{sec:Dynamico_HJ_confirm}), but also that, when the simulations are
extended to long-enough time-scales, an adiabatic profile develops within the
deep atmosphere (\autoref{sec:main_results}). We then explore the robustness of
our results by presenting a series of sensitivity tests, including changes in
the outer and deep atmosphere thermal forcing
(\autoref{sec:robustness_investigate}). Finally, in \autoref{sec:conclusion}, we
provide concluding remarks, including suggestions for future computational
studies of hot Jupiter atmospheres and a discussion about implications for the
evolution of highly irradiated gas giants.
\section{Method}
\label{sec:method}
DYNAMICO is a highly computationally efficient GCM that solves the primitive equation of meteorology ({see \citealt{Vallis17} for a review and \citealt{2014JAtS...71.4621D} for a more detailed discussion of the approach taken in DYNAMICO}) on a sphere \citep{gmd-8-3131-2015}. It is being developed as the next state--of--the art dynamical core for Earth and planetary climate studies at the Laboratoire de Météorologie Dynamique and is publicly available\footnote{DYNAMICO is available at http://forge.ipsl.jussieu.fr/dynamico/wiki, and our hot Jupiter patch available at https://gitlab.erc-atmo.eu/erc-atmo/dynamico\_hj.}. It has recently been used to model the atmosphere of Saturn at high resolution \citep{2018arXiv181101250S}. Here, we present some specificities of DYNAMICO (section \ref{sec:dynamico_NS}) as well as the required modifications we implemented to model hot Jupiter atmospheres (section \ref{sec:newtonian_cooling}).
%The \textit{Primitive Equations of Meteorology} 
\subsection{DYNAMICOs numerical scheme}
\label{sec:dynamico_NS}
\begin{table}[!t]
  \centering
  \def\arraystretch{1.5}
  \begin{tabular}{c|c|c}
    Quantity (units) & Description & Value \\
    \hline \hline
    dt (seconds) & Time-step & 120 \\
    $N_z$ & Number of Pressure Levels & 33 \\
    $d$ & Number of Sub-divisions & 20 \\
    $N\left(^\circ\right)$ & Angular Resolution & $~3.5$\\
    $P_{top}$ (bar) & Pressure at Top  & $7 \times 10^{-3}$  \\
    $P_{bottom}$ (bar) & Pressure at Bottom & 200  \\
    $g$ (m.s${^-2}$) & Gravity & 8.0 \\
    $R_{HJ}$ (m) & HJ Radius  & $10^8$ \\
    $\Omega$ (s$^{-1}$) & HJ Angular Rotation Rate & $2.1 \times 10^{-5}$ \\
    $c_p$ (J.kg$^{-1}$.K$^{-1}$) & Specific Heat & 13226.5 \\
    $\mathcal{R}$ (J.kg$^{-1}$.K$^{-1}$) & Ideal Gas Constant & 3779.0 \\
    $T_{init}  (K) $ & Initial Temperature & 1800 \\
  \end{tabular}
  \caption{Parameters for Low Resolution Simulations}
  \label{tab:lr_params}
\end{table}
Briefly, DYNAMICO takes an energy-conserving Hamiltonian approach to solving the primitive equations. This has been shown to be suitable for modelling hot Jupiter atmospheres \citep{Showman_2008,2012ApJ...750...96R}, although this may not be valid in other planetary atmospheres \citep{2019ApJ...871...56M}. Rather than the traditional latitude-longitude horizontal grid (which presents numerical issues near the poles due to singularities in the coordinate system - see the review of \citealt{WILLIAMSON2007} for more details), DYNAMICO uses a staggered horizontal-icosahedral grid (see \citealt{gmd-7-909-2014} for a discussion of the relative numerical accuracy for this type of grids) for which the number of horizontal cells $N$ is defined by the number of subdivisions $d$ of each edge of the main spherical icosahedral\footnote{\ed Specifically, to generate the grid we start with a sphere that consists of 20 spherical triangles (sharing 12 vertex, i.e. grid, points) then, we subdivide each side of each triangle $d$ times, using the new points to generate a new grid of spherical triangles with $N$ total vertices. These vertices then from the icosahedral grid.  }:
\begin{equation}
  N=10 d^2 + 2.
\end{equation}
As for the vertical grid, DYNAMICO uses a pressure coordinate system whose levels can be defined by the user at runtime. Finally, the boundaries of our simulations are closed and stress-free with zero energy transfer (i.e. the only means on energy injection and removal are the Newtonian cooling profile and the horizontal, numerical, dissipation). {\ed Note that, unlike some other GCM models of hot Jupiters (e.g. \citealt{2009JAtS...66..579S,2013ApJ...770...42L,2019ApJ...883....4S}), we do not include an additional frictional (i.e. Rayleigh) drag scheme at the bottom of our simulation domain, instead relying on the hyperviscosity and impermeable bottom boundary to stabilise the system.  }  \\
As a consequence of the finite difference scheme used in DYNAMICO, artificial numerical dissipation must be introduced in order to stabilise the system against the accumulation of grid-scale numerical noise. This numerical dissipation takes the form of a horizontal hyper-diffusion filter with a fixed hyperviscosity and a dissipation time-scale at the grid scale, labelled $\tau_{dissip}$, which serves to adjust the strength of the  filtering (the longer the dissipation time, the weaker the dissipation). Technically DYNAMICO includes three dissipation timescales, each of which either diffuses scalar, vorticity, or divergence independently. However, for our models, we set all three timescales to the same value. It is important to point out that the hyperviscosity is not a direct equivalent of the physical viscosity of the planetary atmosphere, but can be viewed as a form of increased artificial dissipation that both enhances the stability of the code, and accounts for motions, flows, and turbulences which are unresolved at typical grid scale resolutions. This is known as the large eddy approximation and has long been standard practice in the stellar (e.g. \citealt{2005LRSP....2....1M}) and planetary (e.g \citealt{doi:10.1098/rsta.2008.0268}) atmospheric modelling communities. Because it acts at the grid cell level, the strength of the dissipation is resolution dependent at a fixed $\tau_{dissip}$ (this can be seen in our results in \autoref{fig:resolution_scaling}). \\
In a series of benchmark cases, \citet{10.1111/j.1365-2966.2011.18315.x} (hereafter  \citetalias{10.1111/j.1365-2966.2011.18315.x}) have shown that both spectral and finite-difference based dynamical cores which implement horizontal hyper-diffusion filters can produce differences of the order of tens of percent in the temperature and velocity fields when varying the dissipation strength. We also found such a similar sensitivity in our models: for example, the maximum super-rotating jet speed varies between $3000\,\mathrm{ ms^{-1}}$ and $4500\,\mathrm{ms^{-1}}$ as the dissipation strength is varied. The dissipation strength must thus be carefully calibrated. In the absence of significant constraints on hot Jupiter zonal wind velocities, this was done empirically by minimising unwanted small-scale numerical noise as well as replicating published benchmark results (An alternative, which is especially useful in scenarios where direct or indirect data comparisons are unavailable, is to plot the spectral decomposition of the energy profile and adjust the diffusion such that the energy accumulation on the smallest scales is insignificant). We found that setting $\tau_{dissip}=2500\,\mathrm{s}$ in our low resolution runs leads to benchmark cases in good agreement with the results of, for example \citet{Mayne_2014}, whilst also exhibiting minimal small-scale numerical noise. This is in reasonable agreement with other studies, with our models including a hyper-diffusion of the same order of magnitude as, for example, \citetalias{10.1111/j.1365-2966.2011.18315.x}. Note that, due to differences in the dynamics between those of Saturn and that observed in hot Jupiters, and in particular due to the presence of the strong super-rotating jet, we must use a significantly stronger dissipation to counter grid-scale noise than that used in previous atmospheric studies calculated using DYNAMICO \citep{2018arXiv181101250S}.
%The atmosphere is initialised with an isothermal temperature equals to $1800$K and we use $33$ pressure levels uniformly spaced in $\log\left(P\right)$. 
\subsection{Newtonian cooling} 
\label{sec:newtonian_cooling}
In our simulations of hot Jupiter atmospheres using DYNAMICO, we do not directly model either the incident thermal radiation on the day-side, or the thermal emission on the night-side, of the exoplanet. This would be prohibitively computationally expensive for the long simulations we perform in the present work. Instead we use a simple thermal relaxation scheme to model those effects, with a spatially varying equilibrium temperature profile $T_{eq}$ and a relaxation time-scale $\tau$ that increases with pressure throughout the outer atmosphere. Specifically, this is done by adding a source term to the temperature evolution equation that takes the form: 
\begin{equation}
\frac{\partial T\left(P,\theta,\phi\right)}{\partial t} = - \frac{T\left(P,\theta,\phi\right)-T_{eq}\left(P,\theta,\phi\right)}{\tau\left(P\right)} \, .
\end{equation}
This method, known as Newtonian Cooling has long been applied within the 3D GCM exoplanetary community (i.e. \citet{2002A&A...385..166S}, \citet{Showman_2008}, \citealt{2010ApJ...714.1334R}, \citet{2011ApJ...738...71S}, \citealt{2014GMD.....7.3059M}, \citealt{GUERLET2014110} or \citealt{Mayne_2014}), although it is gradually being replaced by coupling with simplified, but more computationally expensive, radiative transfer schemes (e.g. {\ed \citealt{2009ApJ...699..564S}}, \citealt{2012ApJ...750...96R} or \citealt{2016A&A...595A..36A}) due to its limitations (e.g. it is difficult to use to probe individual emission or absorption features, such as non-equilibrium atmospheric chemistry or stellar activity).\\
The forcing temperature and cooling time-scale we use within our models have their basis in the profiles \citet{2005A&A...436..719I} calculated via a series of 1D radiative transfer models. These models were then parametrised by \citetalias{10.1111/j.1365-2966.2011.18315.x}, who created simplified day-side and night-side profiles. The parametrisation used here is based upon this work, albeit modified in the deep atmosphere since this is the focus of our analysis. As a result, it somewhat resembles a parametrised version of the cooling profile considered by \citet{2013ApJ...770...42L}. \\ 
Specifically, $T_{eq}$ is calculated from the pressure dependent night-side profile ($T_{night}\left(P\right)$) according to the following relation:
\begin{equation}
T_{eq}\left(P,\theta,\phi\right) = T_{night}\left(P\right) + \Delta{T}(P) \cos\left(\theta\right)
\max \left[ 0, \cos (\phi - \pi) \right] \, ,
\end{equation}
where $\Delta T$ is the pressure dependent day-side/night-side temperature difference,
\begin{equation}
\Delta T (P) =\left\{ \begin{array}{ll}
  \Delta T_0 & \textrm{if } P<P_{low} \\
 \Delta T_0 \log (P/P_{low}) & \textrm{if } P_{low}  < P < P_{high} \\
 0 & \textrm{if } P > P_{high}  \end{array}
\right. \, , 
\end{equation}
in which we used $\Delta T_0=600$ K, $P_{low}=0.01$ bar and $P_{high}=10$ bar. The night-side temperature profile $T_{night}$ is parametrised as a series of linear interpolations in $\log(P)$ space between the points
\begin{equation}
\left(\frac{T}{1 \textrm{K}},\frac{P}{1 \textrm{ bar}}\right)=(800,10^{-6}) \textrm{, } (1100,1)  \textrm{ \& }
  (1800,10) \, .
\end{equation}
For $P>10$ bar, we set $T_{eq}=T_{night}=T_{day}=1800$K. \\
Likewise, at pressures smaller than $10$ bar, $\tau$ is linearly interpolated, in $\log(P)$ space, between the points
\begin{equation}
\left(\log\left(\frac{\tau}{1 \textrm{sec}}\right),\frac{P}{1 \textrm{
    bar}}\right)=(2.5,10^{-6}) \textrm{, } (5,1) \textrm{, } (7.5,10) \textrm{ \& }
  (\log(\tau_{220}),220) \, . \label{eq:X}
\end{equation}
For $P>10$ bar, we consider a series of models that lie between two extremes: at one extreme we set $\log(\tau_{220})$ {\ed (which we define as the decimal logarithm of the cooling time-scale $\tau$ at the bottom of our model atmospheres: i.e. at $P=220$ bar)} to infinity, which implies that the deep atmosphere is radiatively inert, with no heating or cooling. As for the other extreme, this involves setting $\log(\tau_{220})=7.5$, which implies that radiative effects do not diminish below  $10$ bar. In \autoref{sec:core_results} we explore results at the first extreme, with no deep radiative dynamics. Then, in \autoref{sec:sensitivity_deep}, we explore the sensitivity of our results to varying this prescription.
%\subsection{Limitations Of The Primitive Equations}
%Whilst evidence points to the primitive equations solved within DYNAMICO being valid within the atmospheres of hot Jupiters (e.g. \citealt{Showman_2008}, \citealt{Rauscher_2010}, and \citetalias{10.1111/j.1365-2966.2011.18315.x} have all used the primitive equations to successfully model hot Jupiter atmospheres, and \citealt{Mayne_2014} explored the relative accuracy of the primitive and full equations, finding slight differences, between models calculated using the full and primitive equations, in the evolution of the deep atmosphere), the same cannot be said for all classes of exoplanets. For example the work of \citet{2019ApJ...871...56M} revealed the presence of significant inaccuracies in atmospheric models of warm, slowly rotating, super Earths and small Neptunes when using the primitive equations of Meteorology. Specifically they find that both the Shallow-Fluid (the assumption that the atmosphere is relatively thin) and Traditional (the assumption that the vertical velocity in the horizontal momentum equation tends to zero) approximations break down, with only Hydrostatic Balance holding true. See, for example, \citealt{2008RvGeo..46.2004G} and  \citealt{2015QJRMS.141.3056T} for additional details about the breakdown of the approximations behind the primitive equations in both Earth's and exoplanetary atmospheres. 
\section{Results}
\label{sec:core_results}
\begin{table}[!t]
  \centering
  \def\arraystretch{1.5}
  \begin{tabular}{>{\centering\arraybackslash}m{10mm}|m{65mm}}
    Model & Description \\
    \hline \hline
    {\it A} & The base low resolution model, in which the deep atmosphere is isothermally initialised \\
    {\it B} & Like model {\it A}, but with the deep atmosphere adiabatically initialised  \\
    {\it C} & Mid Resolution version of model {\it A} ($d = 30$)\\
    {\it D} & High Resolution version of model {\it A} ($d = 40$)\\
    {\it E$\rightarrow$I} & Highly evolved versions of model {\it A}, which have reached a deep adiabat and then had deep isothermal Newtonian cooling introduced at various strengths: For {\it E} $\log(\tau_{220})=7.5$, {\it F} $\log(\tau_{220})=11$, {\it G} $\log(\tau_{220})=15$, {\it H} $\log(\tau_{220})=20$, and {\it I} $\log(\tau_{220})=22.5$     \\
    {\it J} \& {\it K} & Highly evolved versions of model {\it A} which have reached a deep adiabat, and then had their outer atmospheric Newtonian cooling modified to reflect a different surface temperature: 1200K in model {\it J} and 2200K in model {\it K}  \end{tabular}
  \caption{Models discussed in this work}
  \label{tab:all_calcs}
\end{table} 
%\hfill
The default parameters used with our models are outlined in \autoref{tab:lr_params}, with the resultant models, as well as the simulation specific parameters, detailed in \autoref{tab:all_calcs}. \\
In \autoref{sec:main_results}, we use the results of models {\it A} and {\it B} to demonstrate the validity of the work of \citetalias{2017ApJ...841...30T} in the time-dependent, three-dimensional, regime. We next explore the robustness and sensitivity of our results to numerical and external effects in \autoref{sec:robustness_investigate}. Note that, throughout this paper, all times are either given in seconds or in Earth years - specifically one Earth year is exactly 365 days. 
\subsection{Validation of the hot jupiter model} 
\label{sec:Dynamico_HJ_confirm}
\begin{figure*}[ht]
\begin{center}
\begin{subfigure}{0.47\textwidth}
\begin{centering}
\includegraphics[width=0.9\columnwidth]{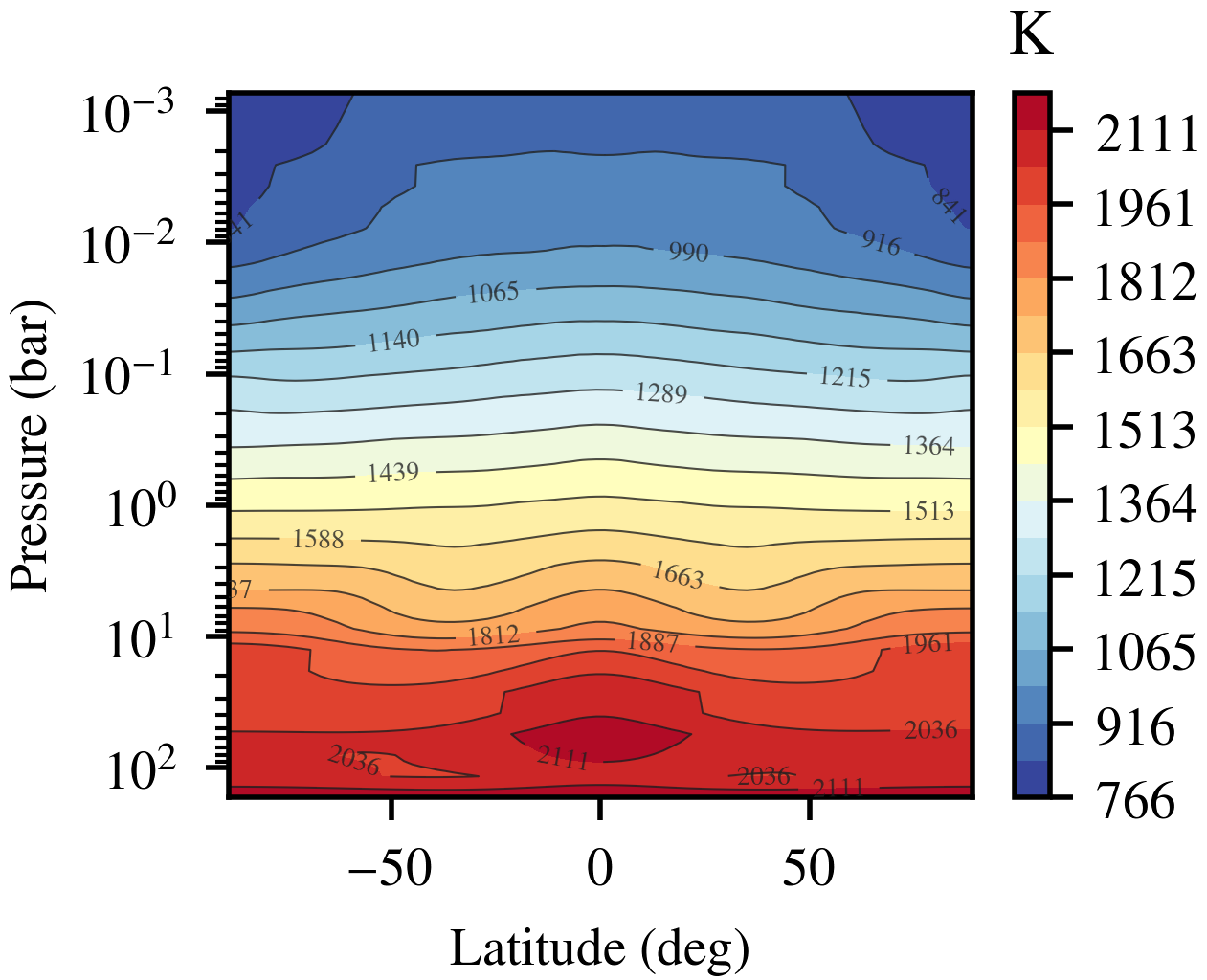}
\caption[]{Temperature Contours  \label{fig:HS_Comp_A} }
\end{centering}
\end{subfigure}
\begin{subfigure}{0.47\textwidth}
\begin{centering}
\includegraphics[width=0.9\columnwidth]{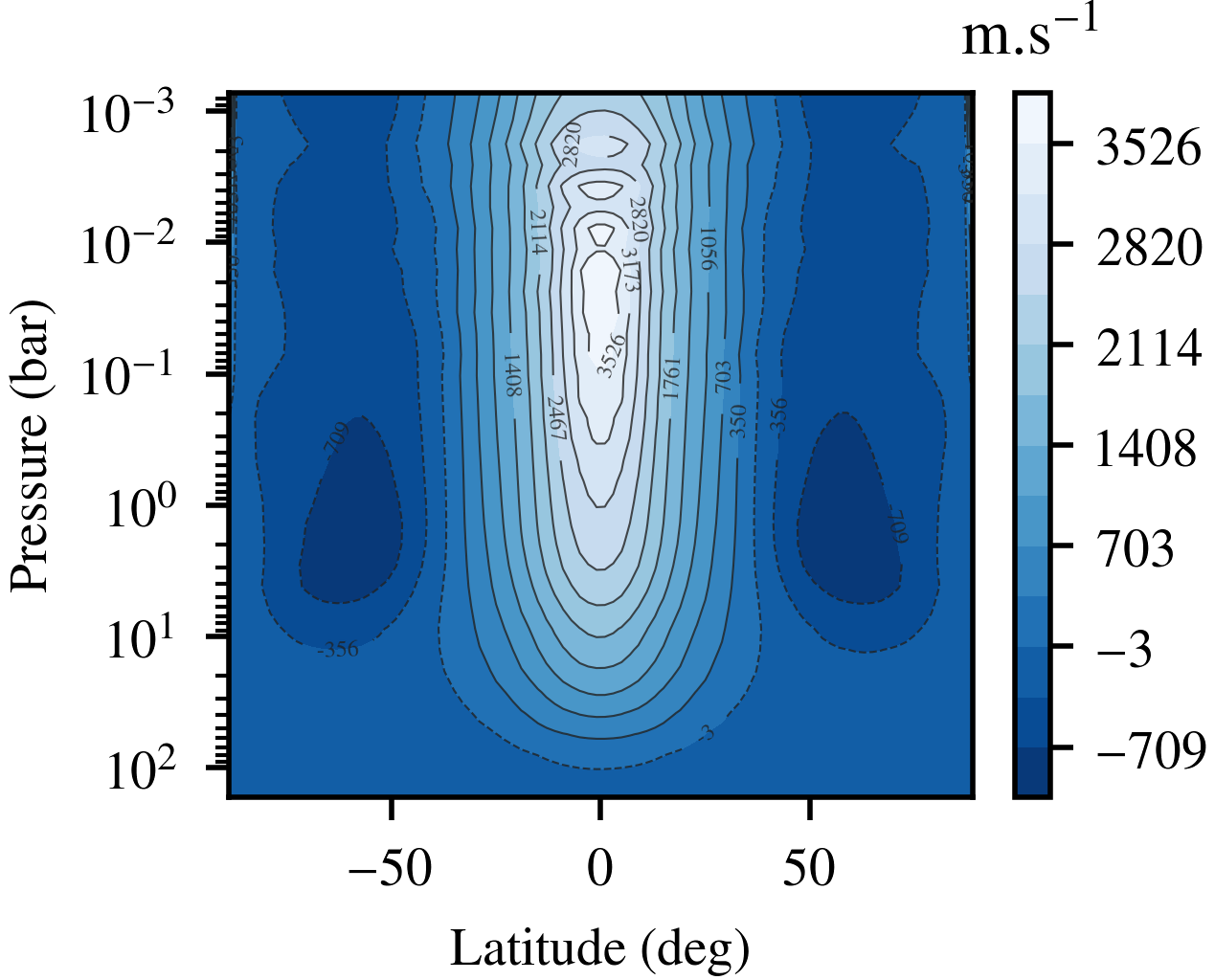}
\caption[]{Zonal Wind Contours   \label{fig:HS_Comp_B} }
\end{centering}
\end{subfigure}
\caption[]{Two plots designed to aid the comparison of the early evolution of model {\it A} with prior studies (e.g. \citet{10.1111/j.1365-2966.2011.18315.x} and \citealt{Mayne_2014}) in-order to benchmark DYNAMICO in the hot Jupiter regime. The figures show the zonally and temporally averaged temperature (a) and zonal wind (b) profiles, both of which are commonly used to benchmark hot Jupiter models. 
}
\label{fig:HS_Comp}
\end{center}
\end{figure*}

We start by exploring the early evolution of model {\it A}, testing how well it agrees with the benchmark calculations of \citetalias{10.1111/j.1365-2966.2011.18315.x}. The model is run for an initial period of $30$ years in order to reach an evolved state before we take averages over the next five $\textrm{years}$ of data. Note that this model was also used to calibrate the horizontal dissipation ($\tau_{dissip}$). In \autoref{fig:HS_Comp}, we show zonally and temporally-averaged plots of the zonal wind and the temperature as a function of both latitude and pressure. \\
We find that the temperature (left panel) is qualitatively similar to that reported by both \citetalias{10.1111/j.1365-2966.2011.18315.x} and \citet{Mayne_2014}. The temperature range we find ($\sim\!\!750\textrm{K}\rightarrow\sim\!\!2150\textrm{K}$) matches their results ($\sim700\rightarrow2000\si{\kelvin}$) to within a 10\% margin of uncertainty. This is satisfactory given the differences between the various set-ups and numerical implementations of the GCMs, as well as the variations that occur when adjusting the length of the temporal averaging window. \\
The zonal wind displays a prominent, eastward, super-rotating equatorial jet that extends from the top of the atmosphere down to approximately $\textrm{10 bar}$ ({\ed Note that, as we continue to run this model for more time, the vertical extent of the jet increases, eventually reaching significantly deeper that 100 bar after 1700 years}). It exhibits a peak wind velocity of $\approx3500\si{\meter\per\second}$, depending upon the averaging window considered, in good agreement with the work of both \citetalias{10.1111/j.1365-2966.2011.18315.x} and \citet{Mayne_2014} who found peak jet speeds on the order of $3500\rightarrow4000\si{\meter\per\second}$.  In the upper atmosphere, it is balanced by counter-rotating (westward) flows at extratropical and polar latitudes. The zonal wind is also directed westwards at all latitudes below $\sim\!\!\textrm{50 bar}$, with this wind also contributing to the flows balancing the large mass and momentum transport of the super-rotating jet. \\ 
The differences we find between our models and the reference models are not unexpected. As discussed by \citetalias{10.1111/j.1365-2966.2011.18315.x}, the jet speed and temperature profile are indeed highly sensitive not only to the numerical scheme adopted by the GCM (i.e. spectral vs finite difference - see their Figure 12) but also to the form and magnitude of horizontal dissipation and Newtonian cooling used. In our models, unlike \citetalias{10.1111/j.1365-2966.2011.18315.x}, we explicitly set our deep ($P>10\textrm{bar}$) cooling to zero, which may explain the enhanced deep temperatures observed in our models, most likely an early manifestation of the deep adiabat we expect to eventually develop.  \\
As noted by other works (e.g. \citealt{2009ApJ...700..887M,2010ApJ...714.1334R,Mayne_2014}), it takes a long time for the the deep atmosphere to reach equilibrium, and the above simulation is by no means an exception: the eastward equatorial jet extends deeper and deeper as time increases, with no sign of stopping by the end of the simulated duration. This long time-scale evolution is explored in detail in the following section.
\subsection{The formation of a deep adiabat} 
\label{sec:main_results}
\begin{figure*}[tbp] %
\begin{centering}
\begin{subfigure}{0.47\textwidth}
\begin{centering}
\includegraphics[width=0.9\columnwidth]{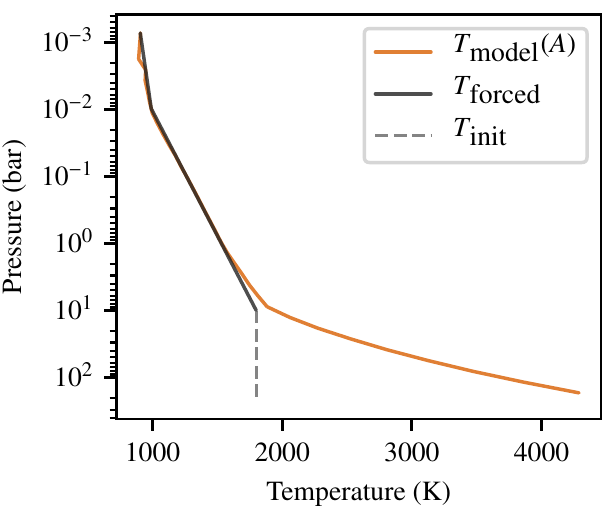}
\caption[]{Isothermal Initialisation   \label{fig:iso_adia_comparison_A} }
\end{centering}
\end{subfigure}
\begin{subfigure}{0.47\textwidth}
\begin{centering}
\includegraphics[width=0.9\columnwidth]{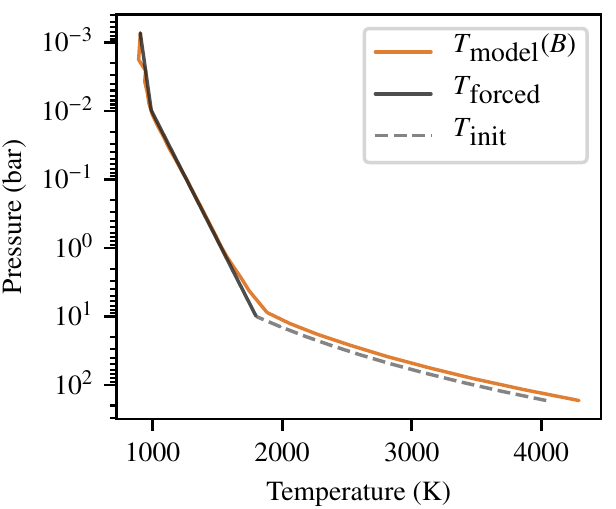}
\caption[]{Adiabatic Initialisation   \label{fig:iso_adia_comparison_B} }
\end{centering}
\end{subfigure}
\caption[Pressure-Temperature profiles for an evolved isothermally initialised case and an adiabatic steady state run. ]{ Equatorially averaged {\ed (i.e. the zonal-mean at the equator)} $T$--$P$ profiles, in orange, for two evolved models that were either isothermally (a) or adiabatically (b) initialised. In both cases there is no forcing below 10 bars (i.e. when $P > \textrm{10 bar}$), and the forcing above this point is plotted in dark grey. {\ed In both cases, the models have been run long enough such that their $T$--$P$ profiles have fully evolved from their initial states}, either isothermal (a) or adiabatic (b) for $P > \textrm{10 bar}$, as shown by the light grey dashed line, {\ed to the same steady state}, a deep adiabat that corresponds to $T_\textrm{surface}=\sim\! 1900\textrm{K}$ - which is $\sim\! 100\textrm{K}$ hotter than the equilibrium temperature at 10 bar.  \label{fig:iso_adia_comparison} }
\end{centering}
\end{figure*}
\begin{figure*}[tbp] %
\begin{centering}
\includegraphics[width=0.9\textwidth]{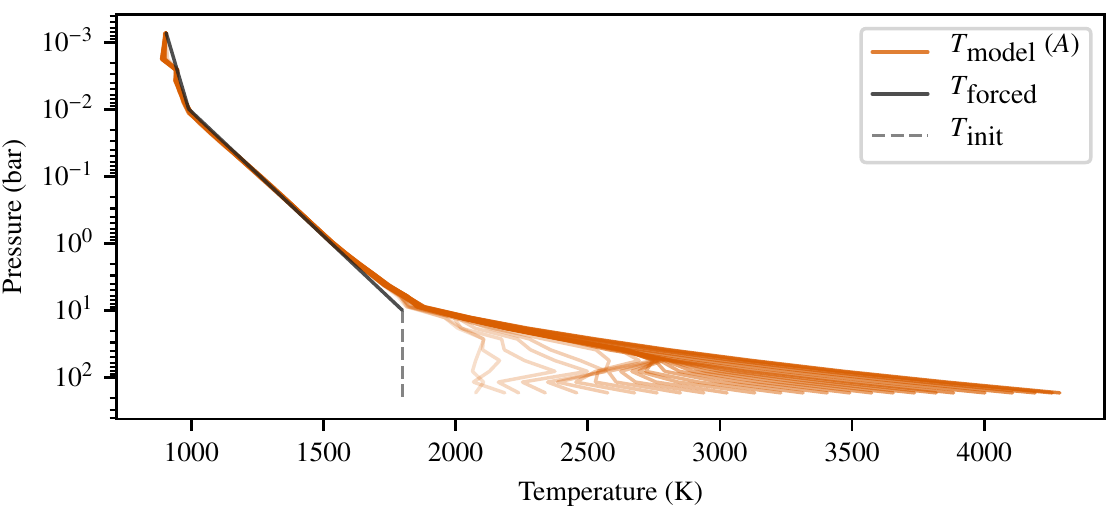}
\caption[Time evolution of the equatorially averaged Temperature-Pressure profile for a low resolution run]{ Time evolution of the equatorially averaged $T$--$P$ profile within model {\it A} covering the $>1500$ (Earth) years of simulation time required for it to reach equilibrium. The light grey dashed line shows the initial temperature profile for $P>10$ bar, whilst the dark grey line shows the forcing profile for $P<10$ bar. The time evolution is represented by the intensity of the lines, with the least evolved (and thus lowest visual intensity) snapshot starting at $t\approx\textrm{30 years}$ followed by later snapshots at increments of approximately $\textrm{60 years}$ \label{fig:time_evo_from_iso}}
\end{centering}
\end{figure*}
\begin{figure*}[hbpt]
\begin{subfigure}[b]{0.5\textwidth}
\begin{centering}
\includegraphics[width=0.95\columnwidth]{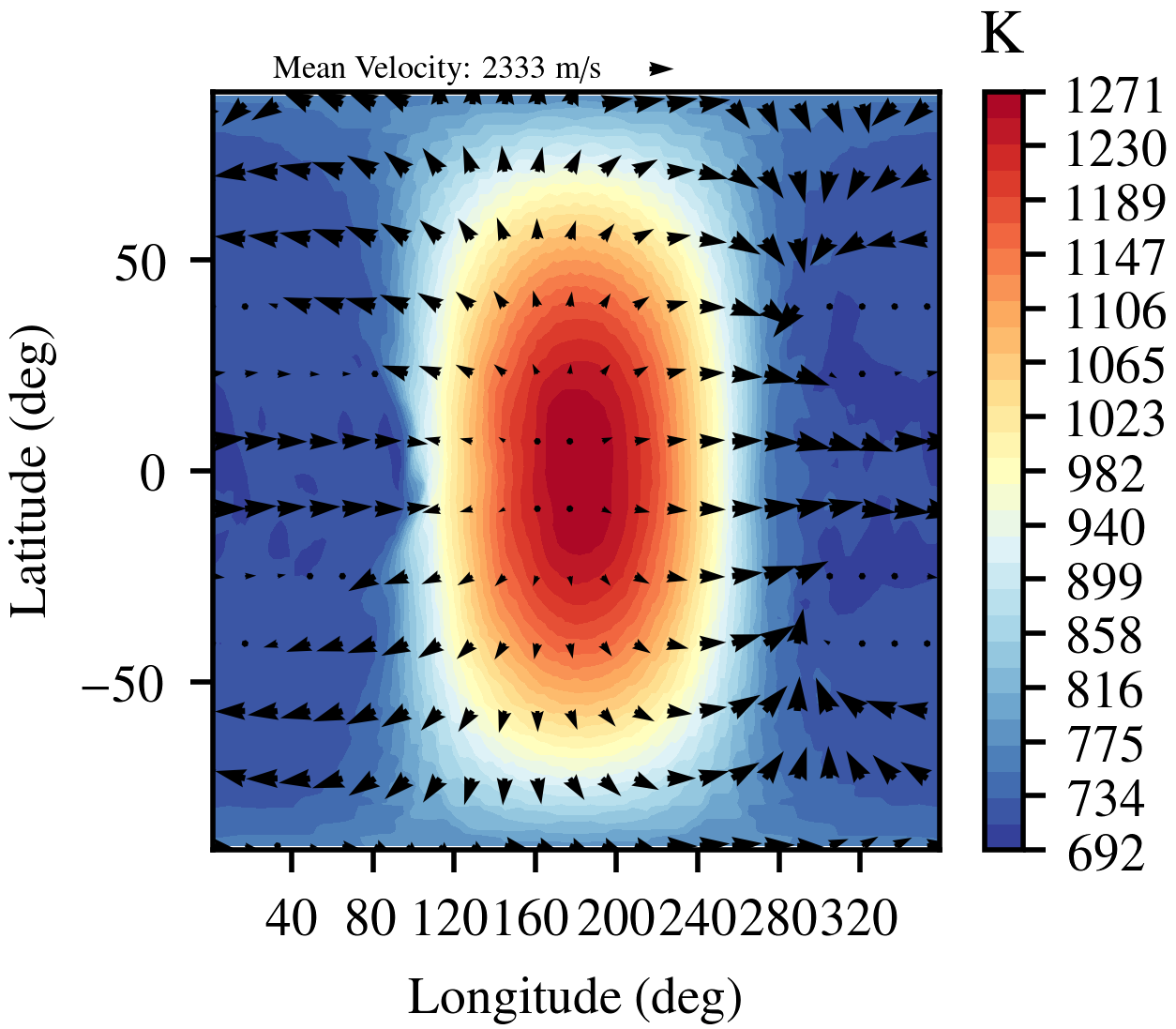}
\caption[]{ {\ed Snapshot of the zonal wind and temperature profile at $P=\textrm{0.72 mbar}$}  \label{fig:temp_wind_snap_A} }
\end{centering}
\end{subfigure}
\hfill
%{}
\begin{subfigure}[b]{0.5\textwidth}
\begin{centering}
\includegraphics[width=0.95\columnwidth]{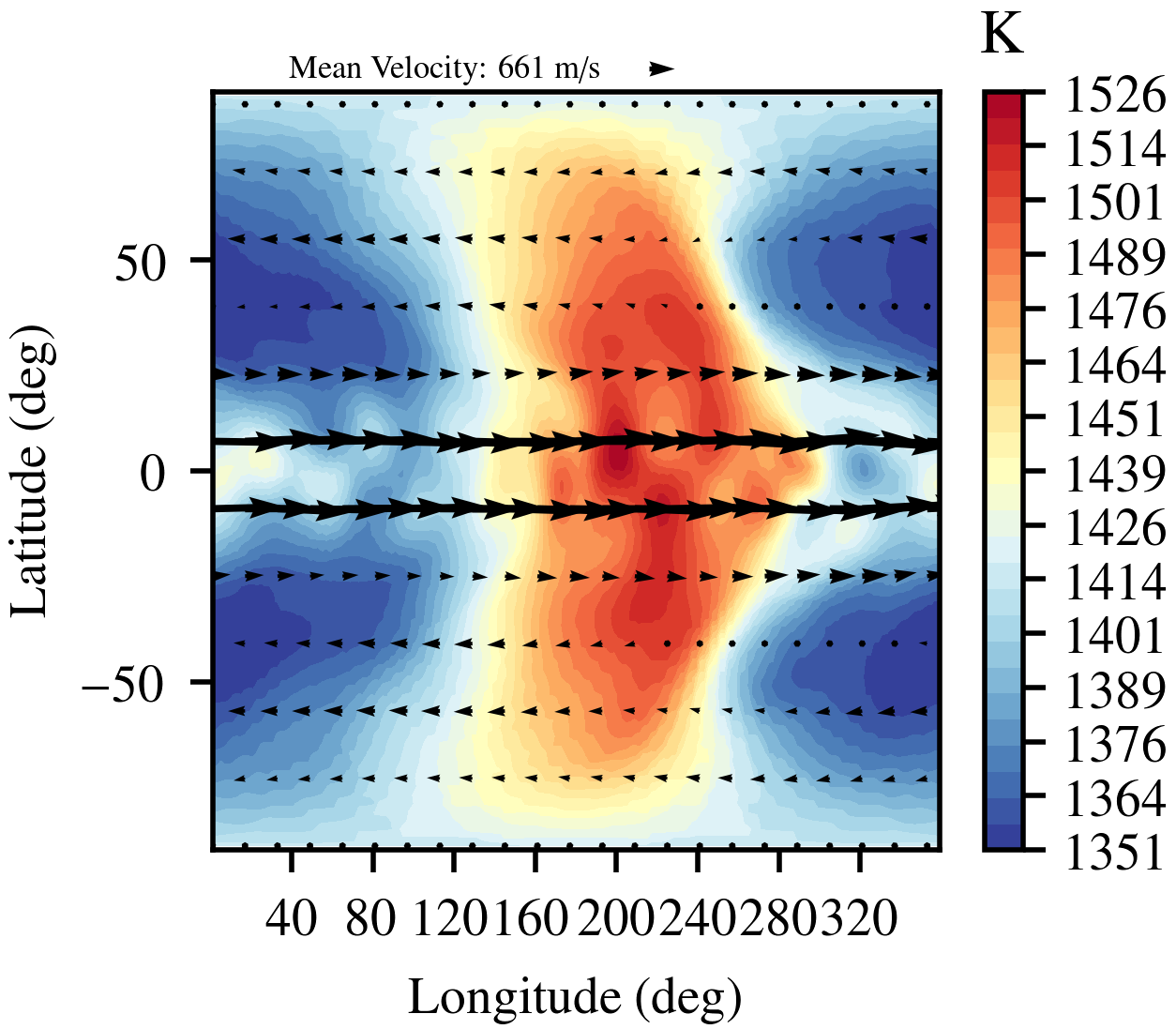}
\caption[]{\ed Snapshot of the zonal wind and temperature profile at $P=\textrm{455 mbar}$  \label{fig:temp_wind_snap_B} }
\end{centering}
\end{subfigure}
\begin{subfigure}[b]{0.5\textwidth}
\begin{centering}
\includegraphics[width=0.95\columnwidth]{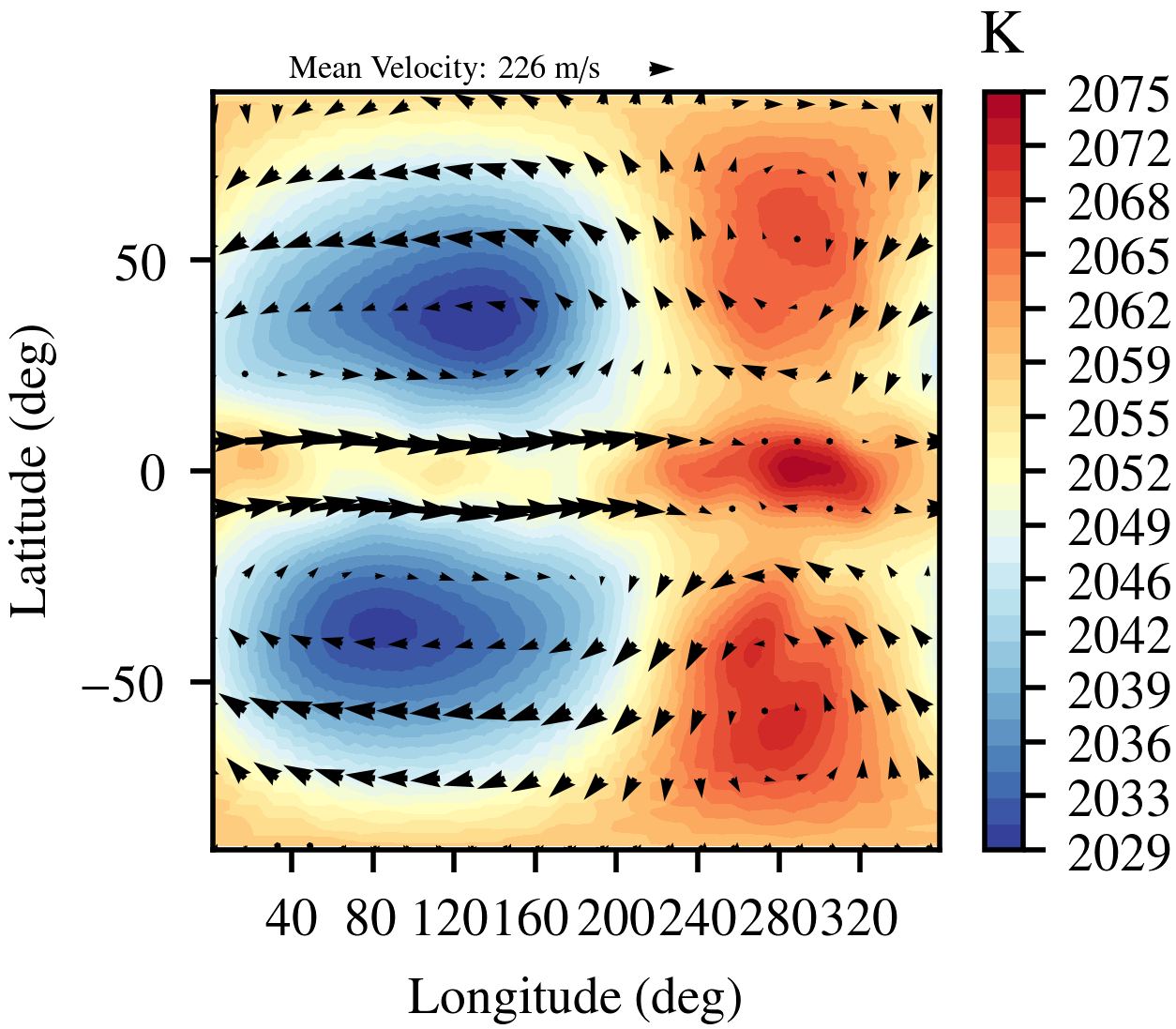}
\caption[]{\ed Snapshot of the  zonal wind and temperature profile at $P=\textrm{12.7 bar}$ \label{fig:temp_wind_snap_C} }
\end{centering}
\end{subfigure}
\hfill
\begin{subfigure}[b]{0.5\textwidth}
\begin{centering}
\includegraphics[width=0.95\columnwidth]{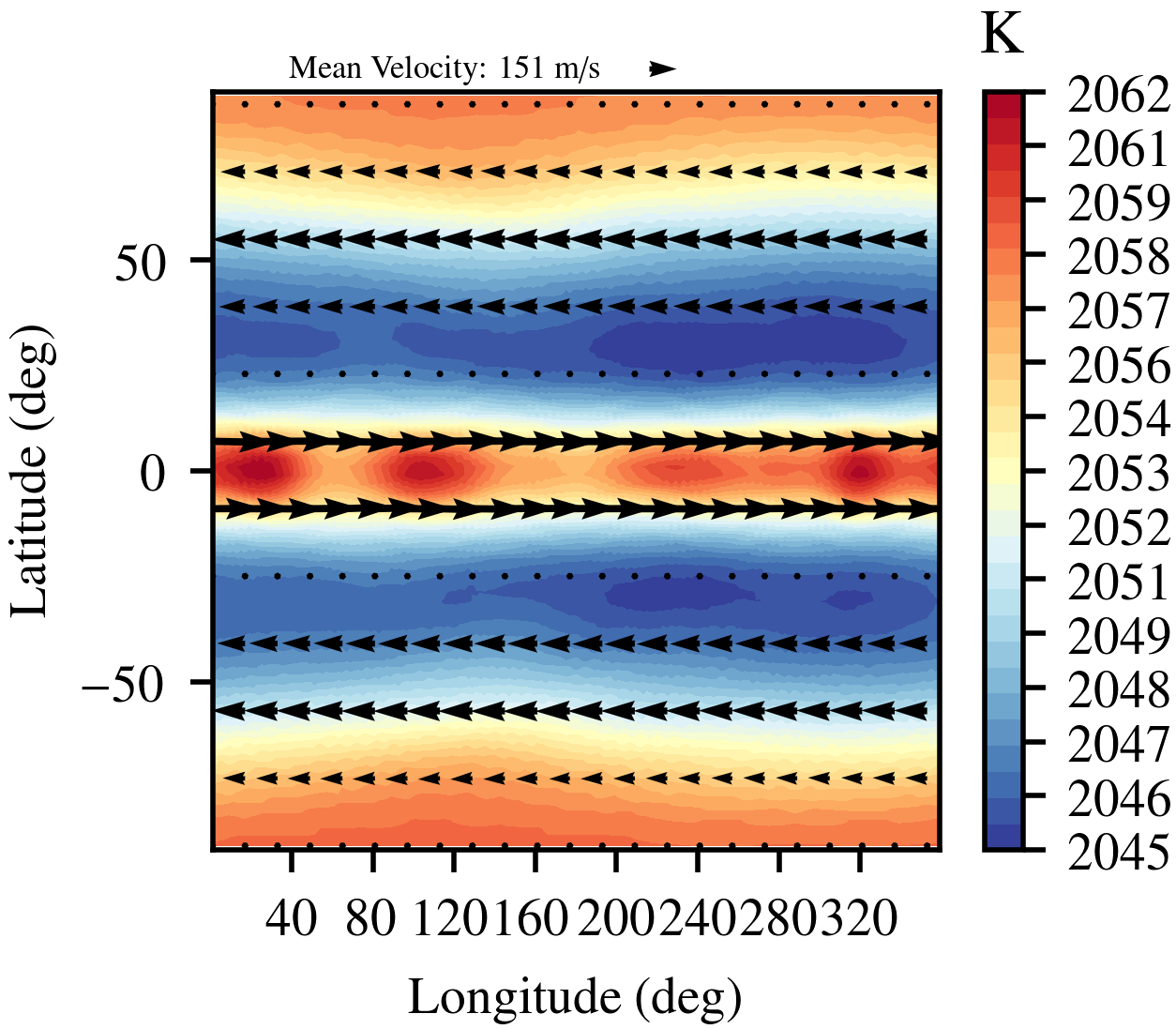}
\caption[]{\ed Time averaged zonal wind and temperature profile at $P=\textrm{12.7  bar}$  \label{fig:temp_wind_snap_D} }
\end{centering}
\end{subfigure}
\caption[]{\ed Zonal wind (arrows) and temperature profile (map) at three different pressures within a fully evolved (i.e. steady-state deep adiabat) model {\it A}. The first three panels ({\it a}, {\it b}, and, {\it c}) show snapshots of the profiles, whilst the last ({\it d}) shows the time-averaged profiles at the same pressure as panel ({\it c}) in order to illustrate the variable nature of the deep atmosphere. Note that the period of the time average was approximately seven and a half years.   \label{fig:temp_wind_snap}}
\end{figure*}

As discussed by \citetalias{2017ApJ...841...30T}, and in \autoref{sec:introduction},  an adiabatic profile in the deep atmosphere (i.e. $P>\sim\!\!1\rightarrow \sim\!\!\textrm{10 bar}$) should be a good representation of the steady state atmosphere. In order to confirm that this is the case, we performed a series of calculations with a radiatively inert deep atmosphere (i.e. no deep heating or cooling, as required by the theory of \citetalias{2017ApJ...841...30T}). \\
We explore this using two models, {\it A} and {\it B}, which only differ in both
 their initial condition and their duration. In model {\it A}, the atmosphere,
 including the deep atmosphere, is initially isothermal with $T$=$1800$K and is
 evolved for more than $1500$ Earth years in order to reach a steady state in its
 $T$--$P$ profile (as shown in \autoref{fig:iso_adia_comparison_A}). As a
 consequence of the long time-scales required for the model to reach equilibrium,
 and the computational cost of such an endeavour, model {\it A} (and {\it B}) is
 run at a relatively low resolution\footnote{ However this does not mean that
   our models have problems conserving angular momentum, they maintain $97.44\%$ of the initial angular momentum after over 1700 years of simulation time (which compares well to other GCMs: \citealt{2014Icar..229..355P}).}. We will investigate the sensitivity of our results to spatial resolution in \autoref{sec:res}.
As for model {\it B}, it is identical to model {\it A} except in the deep atmosphere, where it is initialised with an adiabatic $T$--$P$ profile for $P$>$10$ bar. As a result of this model being initialised close to the expected equilibrium solution, model {\it B} was then run for only $100$ years in order to confirm the stability of the steady-state. In both cases, we find that the simulation time considered is long enough such that the thermodynamic structure of the atmosphere has not changed for multiple advective turnover times $t_{adv} \sim 2 \pi R_{HJ} /u_{\phi}$. \\
\autoref{fig:iso_adia_comparison} shows that both models have evolved to the
same steady state: an outer atmosphere whose $T-P$ profile is dictated by the
Newtonian cooling profile, and a deep adiabat which is slightly hotter
$\left(\sim\!\!1900\textrm{K}\right)$ than the cooling profile at $P=\textrm{10
  bar}$ $\left(1800\textrm{K}\right)$. {\ed  This is reinforced by the latitudinal and longitudinal temperature profile throughout the simulation domain. In \autoref{fig:temp_wind_snap} we plot the zonal wind and temperature profile at three different heights (pressures). Here we can see that, in the outer atmosphere (panels {\it a} and {\it b}) the profile is dominated by the newtonian cooling, with horizontal advection (and the resulting offset hotspot) starting to become significant as we move towards middle pressures. As for the deep atmosphere (snapshot in panel {\it c} and time average in panel {\it d}), here we start to see evidence of both the heating and near-homogenisation of the deep atmosphere. Note that we refer to the atmosphere as nearly homogenised because the temperature fluctuations at, for example, $P=10$ bar are less than $1\%$ of the mean temperature.}\\
Importantly, this convergence to as deep
adiabat not only occurs in the absence of vertical convective mixing (an effect which
is absent from our models, which contain no convective
driving), but also at a significantly lower pressure $\left(P=\textrm{10 bar}\right)$ than the pressure ($\sim$40 bar for HD209458b -
\citealt{2004ApJ...603L..53C}) at which we would expect the atmosphere to become
unstable to convection (and so, in the traditional sense, prone to an adiabatic
profile). \\
Therefore, the characteristic entropy profile of the planet is warmer than the
entropy profiles calculated from standard 1D irradiated models. We will discuss
the implications of this result for the evolution of highly irradiated gas giant
in \autoref{sec:conclusion}. \\
%
%
%This in turn implies a larger radius at a given mass, thus providing a possible solution to the radius inflation, and re-inflation, problem. Additional tests (not shown as part of this work) have confirmed that the pressure at which the $T$--$P$ profile becomes adiabatic is tied to the level at which the effects of irradiation and thermal emission are turned off. This is in agreement with the analysis detailed in \citetalias{2017ApJ...841...30T} and \autoref{sec:introduction}.
In model {\it A}, the steady state described above is very slow to emerge from
an initially isothermal atmosphere. This is illustrated in
\autoref{fig:time_evo_from_iso} which shows the time evolution of the $T$-$P$
profile. It takes more than 500 years of simulation time to stop exhibiting a
temperature inversion in the deep atmosphere, let alone the $>$1500 years
required to reach the same steady state as model {\it B}.

%{\ed This slow evolution of the deep atmosphere can also be observed when we look at the kinetic energy evolution: this reveals a kinetic energy profile that rapidly reaches a maximum (of $\sim 10^{10}\textrm{J}$) as the outer atmosphere spins up, before decreasing to an equilibrium state as the outer atmosphere equilibrates. From then until the end of the simulated time it slowly decreases by around a factor of 2, with this decrease being primarily driven by a decrease in the deep kinetic energy (which itself is driven by a similar decease in the deep atmosphere density as it heats at constant pressure levels). Note that a loss of angular momentum can be ruled out from being the primary driver in the decrease of the kinetic energy, with our models maintaining $97.44\%$ of the initial angular momentum after over 1700 years of simulation time (which compares well to other GCMs: \citealt{2014Icar..229..355P}).    }
As will be further discussed in \autoref{sec:conclusion}, this {\ed slow evolution of the deep adiabat} is probably one of the main reason why this result has not been reported by prior studies of hot Jupiter atmospheres.\\

\begin{figure}[btp] %
\begin{centering}
\includegraphics[width=0.9\columnwidth]{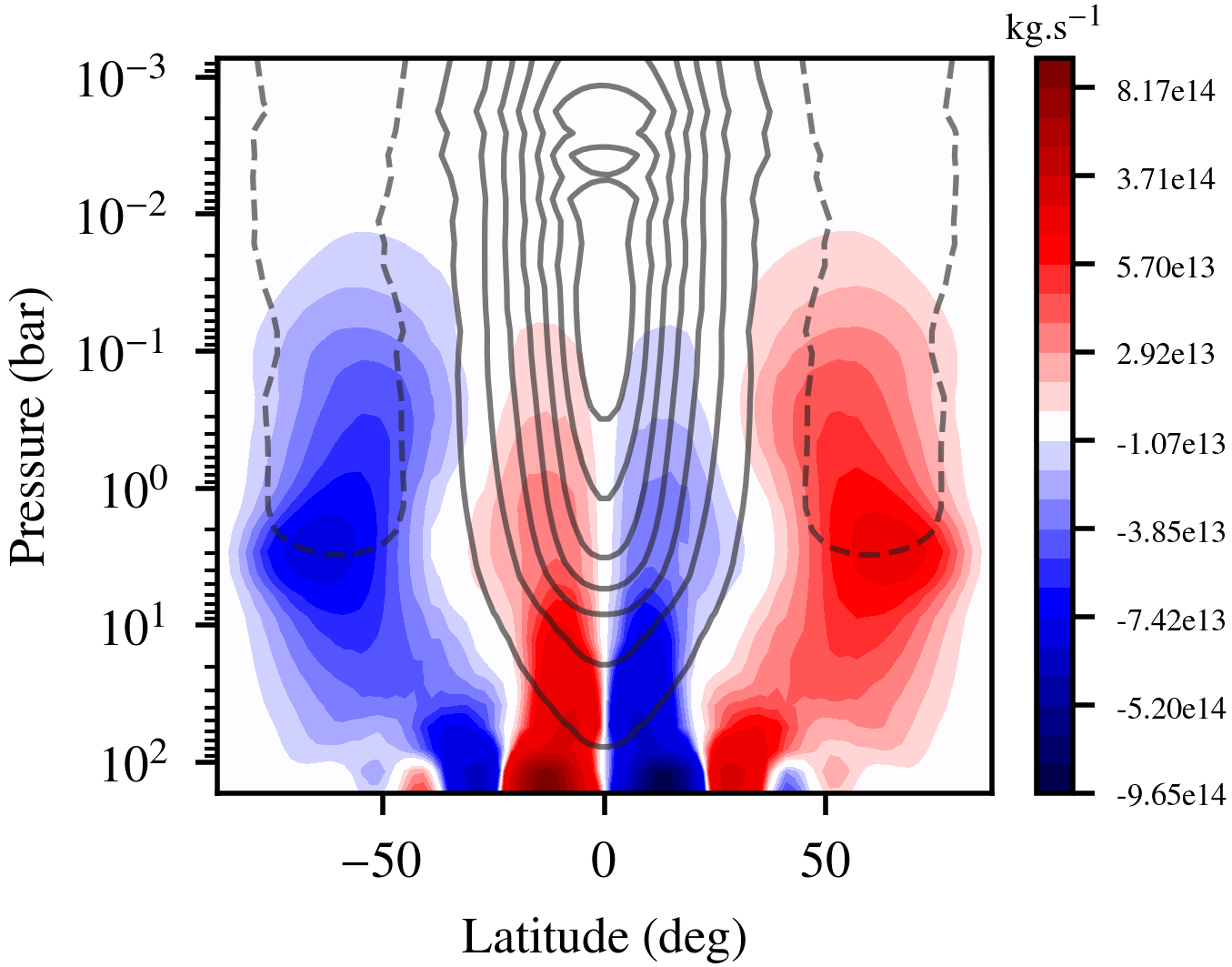}
\caption[Streamfunction showing the vertical and meridional mass flows.]{Zonally and temporally-averaged (over a period of $\approx30$ years) stream-function for model {\it A}. Clockwise circulations on the meridional plane are shown in red and anticlockwise circulations are shown in blue. Additionally the zonally and temporally averaged zonal wind is plotted in black (solid = eastward, dashed = westward).  \label{fig:vertical_streamfuction_mass_flow} }
\end{centering}
\end{figure}

The mechanism advocated by \citetalias{2017ApJ...841...30T} relies on the existence of vertical and latitudinal motions that efficiently redistribute potential temperature. In order to determine their spatial structure, we plot in \autoref{fig:vertical_streamfuction_mass_flow} the zonally and temporally averaged meridional mass-flux stream function and zonal wind velocity for model {\it A}. \\
{\ed Starting with the zonal wind profile (grey lines) we can see evidence for a
  super-rotating jet that extends deep into the atmosphere, with balancing
  counter flows at the poles and near the bottom of the simulation domain. In
  the deep atmosphere, this jet has evolved with the deep adiabat, extending
  towards higher pressures as the developing adiabat (almost) homogenises (and
  hence barotropises) the atmosphere. This barotropisation on long timescales
  seems similar to the drag-free simulation started from a barotropic
zonal wind in \citet{2013ApJ...770...42L}}

%  Evidence for the effect of this barotropisation can be seen near the poles, where an almost Taylor-Proudman (\citealt{1917RSPSA..93...99T,1916RSPSA..92..408P}) velocity profile can be seen (i.e. a velocity profile that is constant with cylindrical radius, or spherical radius in the shallow-atmosphere approximation).   }  \\ 
The meridional mass-flux stream function is defined according to
\begin{equation}
  \Psi = \frac{2\pi R_{HJ}}{g\cos{\theta}}\int_{P_{top}}^{P}u_{\phi}\,dP.
\end{equation}
We find that the meridional (latitudinal and vertical) circulation profile is dominated by four vertically aligned cells extending from the bottom of our simulation atmospheres to well within the thermally and radiatively active region located in the upper atmosphere. These circulation cells lead to the formation of a strong, deep, down-flow at the equator (which can be linked to the high equatorial temperatures in the upper atmosphere), weaker, upper atmosphere, downflows near the poles, and a mass conserving pair of upflows at mid latitudes ($\theta=20^\circ \rightarrow30^\circ$). The meridional circulation not only leads to the vertical transport of potential temperature (as high potential temperature fluid parcels from the outer atmosphere are mixed with their `cooler' deep atmosphere counterparts), but also to the {\ed almost complete} latitudinal homogenisation of the deep atmosphere {\ed (with only small temperature variations remaining)}. In a fully radiative model, these circulations would also mix the outer atmosphere, leading to the equilibrium temperature profiles we instead impose via Newtonian cooling (see, for example, \citealt{2018A&A...612A.105D,2018ApJ...855L..31D} for more details about the 3D mixing in radiative atmospheres). \\
{\ed Note that the vertical extent the zonal wind, and the structure of the lowest cells in the mass-flux stream function,  appear to be affected by the bottom boundary, suggesting that they extend deeper into the atmosphere. Whilst this is interesting and important, it should not affect the final state our P-T profiles reach, but does suggest that models of hot-Jupiters should be run to higher pressures to fully capture the irradiation driven deep flow dynamics. \\}

The primary driver of the latitudinal homogenisation are fluctuations in the meridional circulation profile, which are visible within individual profile snapshots, but are averaged out when we take a temporal average. This includes contributions from spatially small-scale
velocity fluctuations at the interface of the large-scale meridional cells. Evidence for
these effects can be seen in snapshots of the zonal and meridional flows, in an RMS analysis
of the zonal velocity, {\ed and of course in deep temperature profile that these advective motions drive}.  The first
reveals complex dynamics, such as zonally-asymmetric and temporally variable
flows, that are hidden when looking at the temporal average, but which mask the
net flows when looking at a snapshot of the circulation. The second reveals spatial and temporal fluctuations on the
order of $5\rightarrow10\si{\meter\per\second}$ in the deep atmosphere. {\ed Finally the third (as plotted in panels {\it c} and {\it d} of \autoref{fig:temp_wind_snap}, which show snapshots or the time average of the zonal wind and temperature profile, respectively) reveals small scale temperature and wind fluctuations, which are likely associated with the deep atmosphere mixing, that are lost when looking at the average, steady, state.  } \\
However, a more detailed analysis of the dynamics of this homogenisation, {\ed as well as the exact nature of the driving flows and dynamics}, is beyond the scope {\ed of this paper. Although interesting in its own right, the mechanism by which the circulation is set up in the deep atmosphere of our isothermally initialised simulations might not be relevant to the actual physical mechanism happening in hot Jupiters with hot, deep, atmospheres.}
%of this paper since in the present study we are primarily interested in the {\it average steady-state flow}. \\
%{\bf SF: I would remove the following that is in bold face. It speaks about the driving of the zonal jet which is not relevant for the mixing that is the focus of this paragraph (we may want to move that to section 3.1), and it also describes that it consists of a strong super--rotating...., which has been said already in section 3.1: These zonally-asymmetric flows are of particular interest since they are known to play in important role in the atmospheric circulations of both the Earth (e.g. \citealt{doi:10.1146/annurev.earth.34.031405.125144}), and of solar-system gas giants (e.g. \citealt{2018arXiv181101250S}). For example, as discussed by \citet{2018arXiv181101250S},  \citet{2011ApJ...738...71S}, and \citet{2014ApJ...793..141T}, zonally-asymmetric effects help to drive the zonal jet (black contours in \autoref{fig:vertical_streamfuction_mass_flow}) and mean meridional circulations. The structure of this zonal jet us well known (e.g. \citep{2011ApJ...738...71S,2014ApJ...793..141T}) and in our models (\autoref{fig:vertical_streamfuction_mass_flow}) consists of a strong super-rotating equatorial jet flanked by two weaker super--rotating jets.}\\
%
\begin{figure*}[btp] %
\begin{centering}
\begin{subfigure}{0.47\textwidth}
\begin{centering}
\includegraphics[width=0.9\columnwidth]{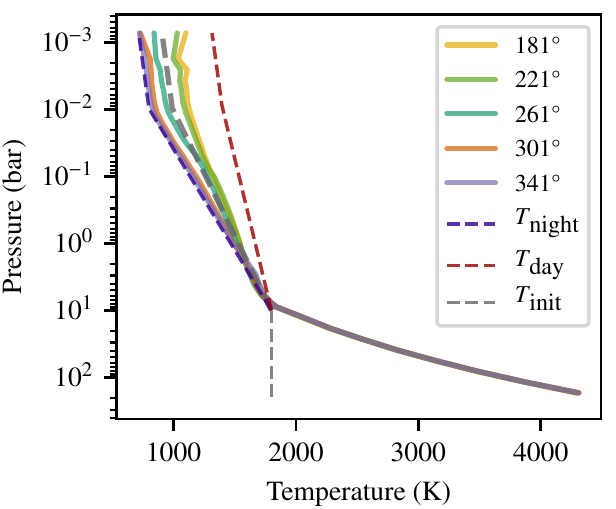}
\caption{Longitudinal Variations \label{fig:T_P_longitude_A}}
\end{centering}
\end{subfigure}
\begin{subfigure}{0.47\textwidth}
\begin{centering}
\includegraphics[width=0.9\columnwidth]{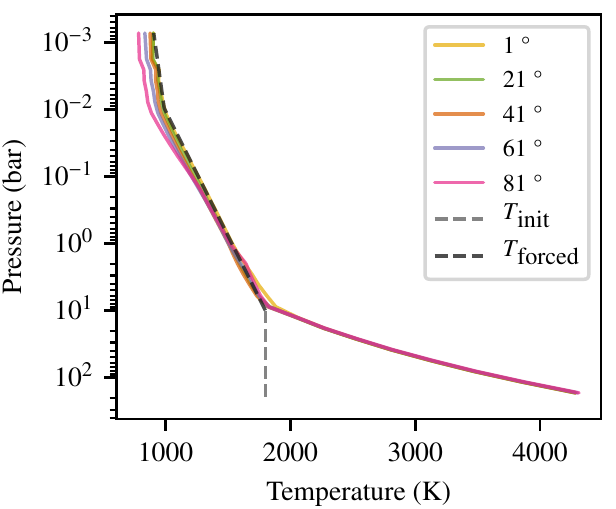}
\caption{Latitudinal Variations \label{fig:T_P_longitude_B}}
\end{centering}
\end{subfigure}
\caption[Change in temperature-pressure profile with longitude.]{Snapshots of the latitudinally (a) or longitudinally (b) averaged T-P profile within model {\it A} at various longitudes or latitudes respectively. In each plot, the solid lines represent the various T-P profiles considered. {\ed At low pressures ($P<10$ bar)}, the dashed lines represent either the days-side, night-side and equilibrium profiles (in the longitudinal plot, a), or the reference Newtonian cooling (in the latitudinal plot, b), {\ed whereas at high pressures ($P>10$ bar), the dashed lines (light grey) represent the initial state of the atmosphere}. \label{fig:T_P_longitude} } %Using these plots, we can see that the deep atmosphere is latitudinally and longitudinally homogenised onto a deep, hot, adiabat. 
\end{centering}
\end{figure*}
% here
As a consequence of both the meridional circulations described above, and the zonal flows that form as a response to the strong day-side/night-side temperature differential, the
deep atmosphere $T$--$P$ profile is independent of both longitude
(\autoref{fig:T_P_longitude_A}) and latitude (\autoref{fig:T_P_longitude_B}). Only in the upper atmosphere ($P<10$ bar) do the temperature profiles
start to deviate from one another, reflecting the zonally and
latitudinally varying Newtonian forcing.  Taken together, the two
panels of \autoref{fig:T_P_longitude} confirm that the latitudinal and vertical
steady-state circulation, the super-rotating eastward jet, and any zonally-asymmetric flows
act to advect potential temperature throughout the deep atmosphere, leading at depth to
the formation of a hot adiabat without the need for any
convective motions.

%% Further, we note that, as expected from
%% previous studies, the amplitudes of the temperature variations are smaller than
%% the imposed day-side/night-side temperature contrast due to the rapid advection
%% of temperature by the super-rotating jet. Further, the variation of the
%% temperature profile with latitude remains modest, reflecting the smaller
%% equator--to--pole temperature gradient in the forcing.

\subsection{Robustness of the results} 
\label{sec:robustness_investigate}

Having confirmed that a deep adiabatic temperature profile connecting with the outer atmospheric temperature profile at $P=\textrm{10 bar}$ is a good representation of the steady state within our hot Jupiter model atmospheres, we now explore the robustness of this result.

\subsubsection{Sensitivity To Changes In The Horizontal Resolution} \label{sec:res}

\begin{figure}[tbp] %
\begin{centering}
\includegraphics[width=0.9\columnwidth]{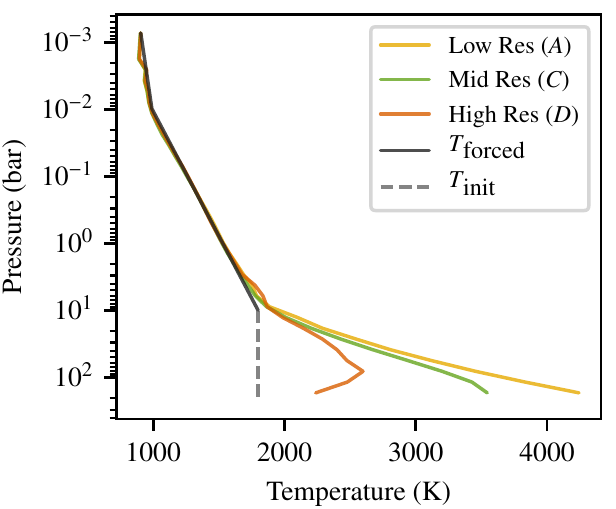}
\caption[Change in temperature-pressure profile with resolution at fixed dissipation.]{Equatorially averaged T-P profile snapshots for three initially isothermal (see grey dashed line in the deep atmosphere) models run with the same dissipation time ($t_{\mathrm{dissip}}=2500\textrm{s}$), vertical resolution, and Newtonian cooling profile (dark grey), but different horizontal resolutions (Models {\it A} (yellow), {\it C} (light green), and {\it D} (orange)).\label{fig:resolution_scaling} }
\end{centering}
\end{figure}

We start our exploration of the robustness of our results by confirming that the
eventual convergence of the deep atmosphere on to a deep adiabat appears resolution independent.\\
\autoref{fig:resolution_scaling} shows the $T$--$P$ profiles obtained for three models at the same time ($t\approx1800$ years) but with different resolutions (our `base' resolution model, {\it A}, a `mid-res' model, {\it C}, and  a `high-res' model, {\it D}). The mid resolution model ({\it C}) has almost reached the exact same equilibrium adiabatic profile as the low resolution case ({\it A}): comparing this with the time-evolution of model {\it A} (\autoref{fig:time_evo_from_iso}) confirms that they are both on the path to the same equilibrium state, and that a significant amount of computational time would be required to reach it. This becomes even clearer when we look at a high resolution model ({\it D}). Here we find that, despite the long time-scale of the computation, the deep atmosphere still exhibits  a temperature inversion, suggesting, in comparison to \autoref{fig:time_evo_from_iso}, that the model has a long way to go until it reaches the same, deep adiabat, equilibrium. \\
In general, we have found the better the resolution the more slowly the atmosphere temperature profile evolves
towards the adiabatic steady state solution. This stems most likely from
the fact that horizontal numerical dissipation, on a fixed dissipation
time-scale, decreases with increasing resolution. Note that we kept the horizontal dissipation timescale constant due to both the computational expense of the parameter study required to set the correct dissipation at each resolution, and the numerical dissipation independence of the steady-state in the deep atmosphere. \\
%This means we might not be corectly accounting for grid-scale noise in hour
%high resolution models. An effect which might affect the atmospheric
%circulations in such a way as to reduce horizontal mixing in the deep
%atmosphere.
Evidence for the impact of the small-scale flows on this slow evolution can be
seen in the temporal and spatial RMS profiles of the zonal flows, which reveal
that, as we increase the resolution by a factor 2, the magnitude of the
small-scale velocity fluctuations decreases by roughly the same factor. These results are in agreement with the effect of changing the numerical dissipation timescale
($\tau_{\textrm{dissip}}$) at a fixed resolution, where longer timescales
also slow down the circulation, thereby increasing the time required to reach a
steady $T$-$P$ profile in the deep atmosphere (not shown). Despite these
numerical limitations, it remains clear that the, the presence, and strength, of any numerical dissipation does not affect the steady state solutions of the
simulation, which remains as an adiabatic P-T profile in the deep atmosphere.

\subsubsection{Sensitivity to changes in the upper atmosphere forcing function}

\begin{figure}[tbp] %
\begin{centering}
\includegraphics[width=0.9\columnwidth]{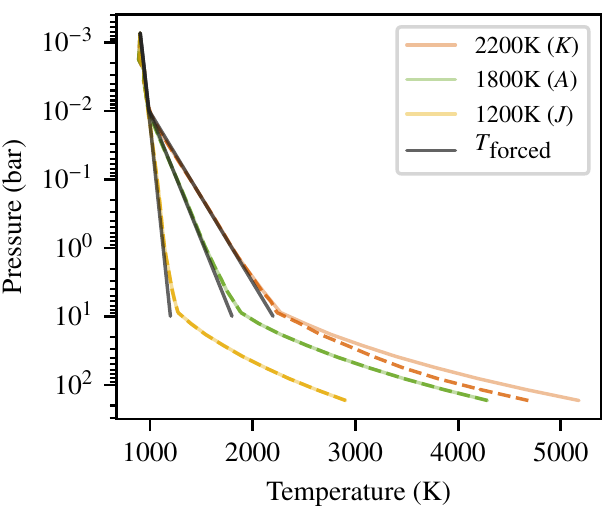}
\caption[Stability of the deep adiabat against low pressure temperature changes]{Equatorially averaged $T$--$P$ profiles for three models: {\it A} (green), {\it J} (yellow) and {\it K} (orange). The orange ({\it K}) and yellow ({\it J}) models have had their outer atmosphere cooling modified such that $T_{\textrm{eq}}=2200$~K or $1200$~K, respectively. The solid lines represent the equilibrium $T$--$P$ profiles whilst the dashed lines represent the $T$--$P$ profiles $200$ years after the outer atmospheres forcing was adjusted (shown in dark grey for each model). Note that, after 200 years of `modified' evolution, only the $2200$K model has not reached equilibrium. } 
\label{fig:force_to_adiabat_tests}
\end{centering}
\end{figure}

We next explore how the deep adiabat responds to changes in the outer atmosphere irradiation and thermal emission (via the imposed Newtonian cooling). The aim is not only to test the robustness of the deep adiabat, but also to explore the response of the adiabat to changes in the atmospheric state. As part of this study, the two scenarios we consider were initialised using the evolved adiabatic profile obtained in model A, but with a modified outer atmosphere cooling profile such that $T_{night}=1200$K (model {\it J}) or $T_{night}=2200$K (model {\it K}). \autoref{fig:force_to_adiabat_tests} shows the equilibrium $T$--$P$ profiles (solid lines) as well as snapshots of the $T$--$P$ profiles after only $200$ years of `modified' evolution (dashed lines). It also includes a plot of model {\it A} to aid comparison.  \\
Model {\it J} evolves in less than 200 years towards a new steady state profile that corresponds to the modified cooling profile. The deep adiabat reconnects with the outer atmospheric profile at $P=\textrm{10 bar}$ and $\sim\!\textrm{1250K}$ (in agreement with the relative offset found in our $1800\textrm{K}$ models, {\it A} and {\it B}). The meridional mass circulation (not shown) displays evidence for the same qualitative flows driving the vertical advection of potential temperature as models {\it A} and {\it B}. However it also shows signs that it is still evolving, suggesting that the steady state meridional circulation takes longer to establish than the vertical temperature profile. \\
In model {\it K}, we find that, $200$ years after modifying the outer atmospheres cooling profile, the deep atmosphere has not yet reached a steady state. In fact it takes approximately $1000$ years of evolution for it to reach equilibrium, which we show as a solid line in \autoref{fig:force_to_adiabat_tests}. This confirms that, model {\it K}, although slow to evolve relative to the cooling case (model {\it J}), does eventually settle onto a deep, equilibrium, adiabat. Additionally, this adjustment occurs significantly faster than the equivalent evolution of a deep adiabat from an isothermal start.   

Based on the results of this section, we conclude that it is faster for the deep atmosphere to cool than to warm when it evolves toward its adiabatic temperature profile. In order to understand this time-scale ordering, we have to note that the only way for the simulation to inject or extract energy is through the fast Newtonian forcing of the upper atmosphere and also that the thermal heat content of the deep atmosphere is significantly larger than that of the outer layers. The deep ($_{d}$) and upper ($_{u}$) atmospheres are connected by the advection of potential temperature that we will rewrite in a conservative form as an enthalpy flux: $\rho c_p T u$ and we simplify the process to two steps between the two reservoirs (assuming they have similar volumes): injection/extraction by enthalpy flux and Newtonian forcing in the upper atmosphere. 
\begin{itemize}
    \item In the case of cooling, the deep atmosphere contains too much energy and needs to evacuate it. It will setup a circulation to evacuate this extra-energy to the upper layers with an enthalpy flux that would lead to an upper energy content set by $\rho_u c_v T_u \sim \rho_u c_v T_{u,init}+\rho_d c_v (T_{d,init}-T_{d,eq})$ if we ignore first Newtonian cooling. $T_u$ would then be very large essentially because of the density difference between the upper and lower atmosphere. The Newtonian forcing term proportional to $-(T_u - T_{u,eq})/\tau$ is then very large and can efficiently remove the energy from the system. 
    \item In the case of heating, the deep atmosphere does not contain enough energy and needs an injection from the upper layers. This injection is coming from the Newtonian forcing and can at first only inject $\rho_u c_v (T_{u,eq} - T_{u,init})$ in the system. The enthalpy flux will then lead to an energy content in the deep atmosphere given by $\rho_d c_v T_d \sim \rho_d c_v T_{d,init}+\rho_u c_v (T_{u,eq}-T_{u,init})$ if we assume that all the extra-energy is pumped by the deep atmosphere. Because of the density difference and the limited variations in the temperature caused by the forcing, the temperature change in the deep atmosphere is small and will require more injection from the upper layers to reach equilibrium. However, even in the most favourable scenario in which all the extra energy is transferred, the Newtonian forcing cannot exceed $-(T_{u,init} - T_{u,eq})/\tau$ which explains why it will take a much longer time to heat the deep atmosphere than to cool it.
\end{itemize}

\subsubsection{Sensitivity to the addition of newtonian cooling to the deep atmosphere}
\label{sec:sensitivity_deep}

\begin{figure}[tbp] %
\begin{centering}
\includegraphics[width=0.9\columnwidth]{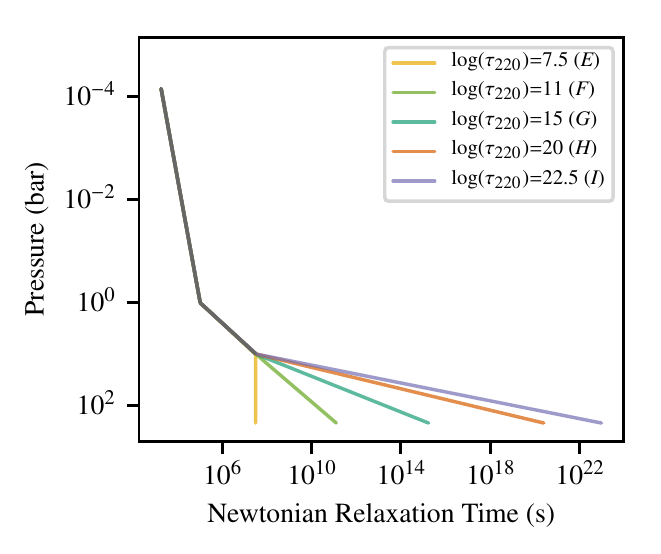}
\caption[Tau profiles used as part of the Newtonian cooling for the
  runs shown in \ref{fig:deep_tau_stability}]{Newtonian cooling
  relaxation time-scale profiles used in the models shown in
  \autoref{fig:deep_tau_stability}. Note that a smaller value of $\tau$ means more rapid forcing towards the imposed cooling profile {\ed (which in all cases is isothermal in the deep atmosphere, where $P>10$ bar)}, and that the relaxation profiles are identical for $P<\textrm{10 bar}$ (grey line). \label{fig:deep_tau_stability_profiles}
}
\end{centering}
\end{figure}

\begin{figure}[tbp] %
\begin{centering}
\includegraphics[width=0.9\columnwidth]{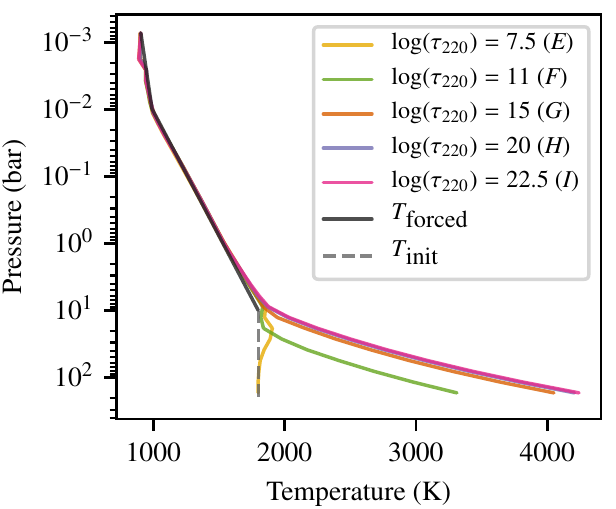} 
\caption[Stability of the deep adiabat against forcing to an Isothermal state.]{Snapshots of the $T$--$P$ profile for five, initially adiabatic simulations (coloured lines - based on model {\it B}, and with the same outer atmosphere cooling profile (dark grey)) which are then forced to a deep isothermal profile (grey dashed line) with varying $\log(\tau_{\textrm{220}})$ (\autoref{eq:X}). \label{fig:deep_tau_stability} }
\end{centering}
\end{figure}

\begin{figure*}[tbp] %
\begin{centering}
\begin{subfigure}{0.47\textwidth}
\begin{centering}
\includegraphics[width=0.9\columnwidth]{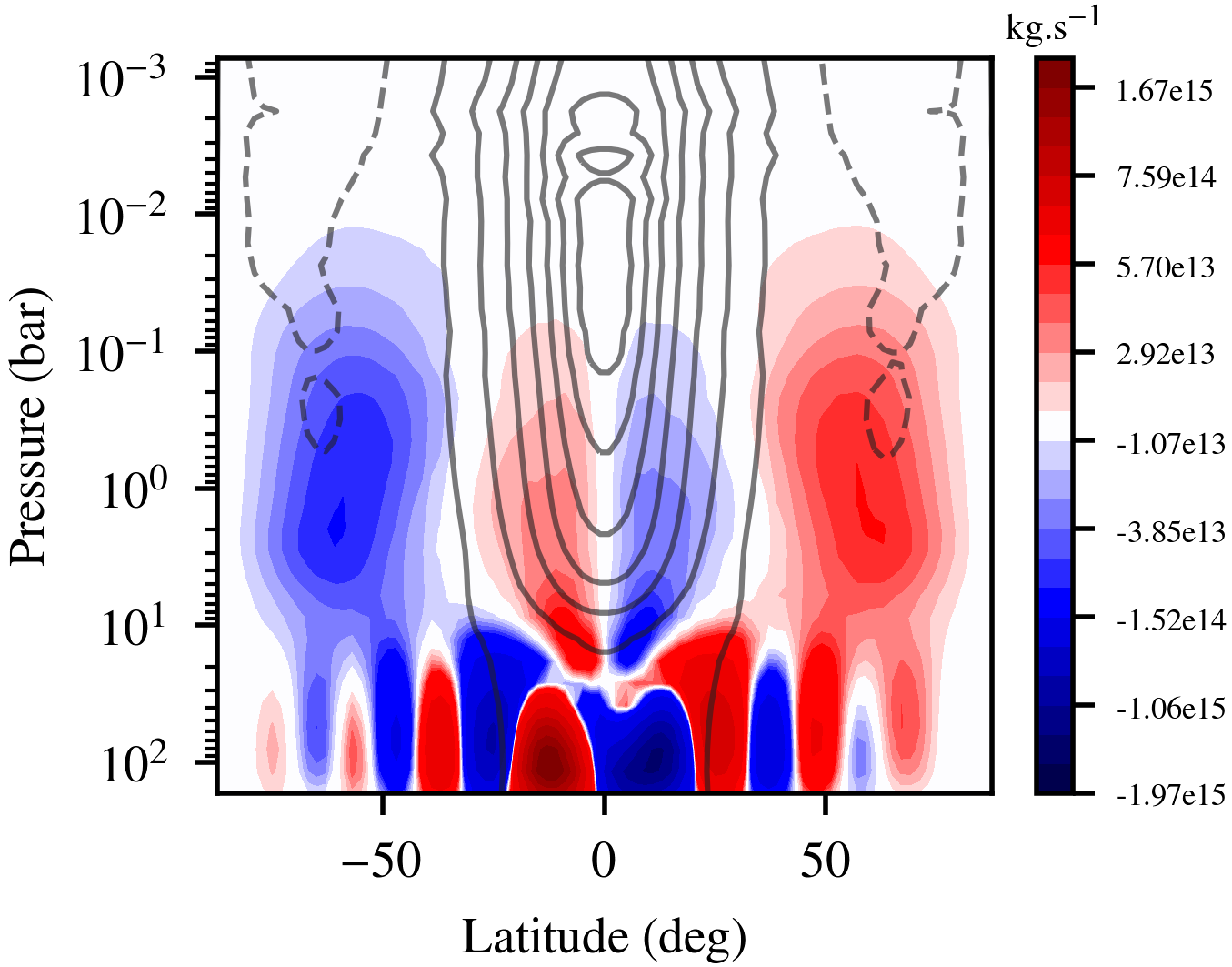} 
\caption[]{$\log(\tau_{220})=15$  \label{fig:deep_tau_forcing_streams_A} }
\end{centering}
\end{subfigure}
\begin{subfigure}{0.47\textwidth}
\begin{centering}
\includegraphics[width=0.9\columnwidth]{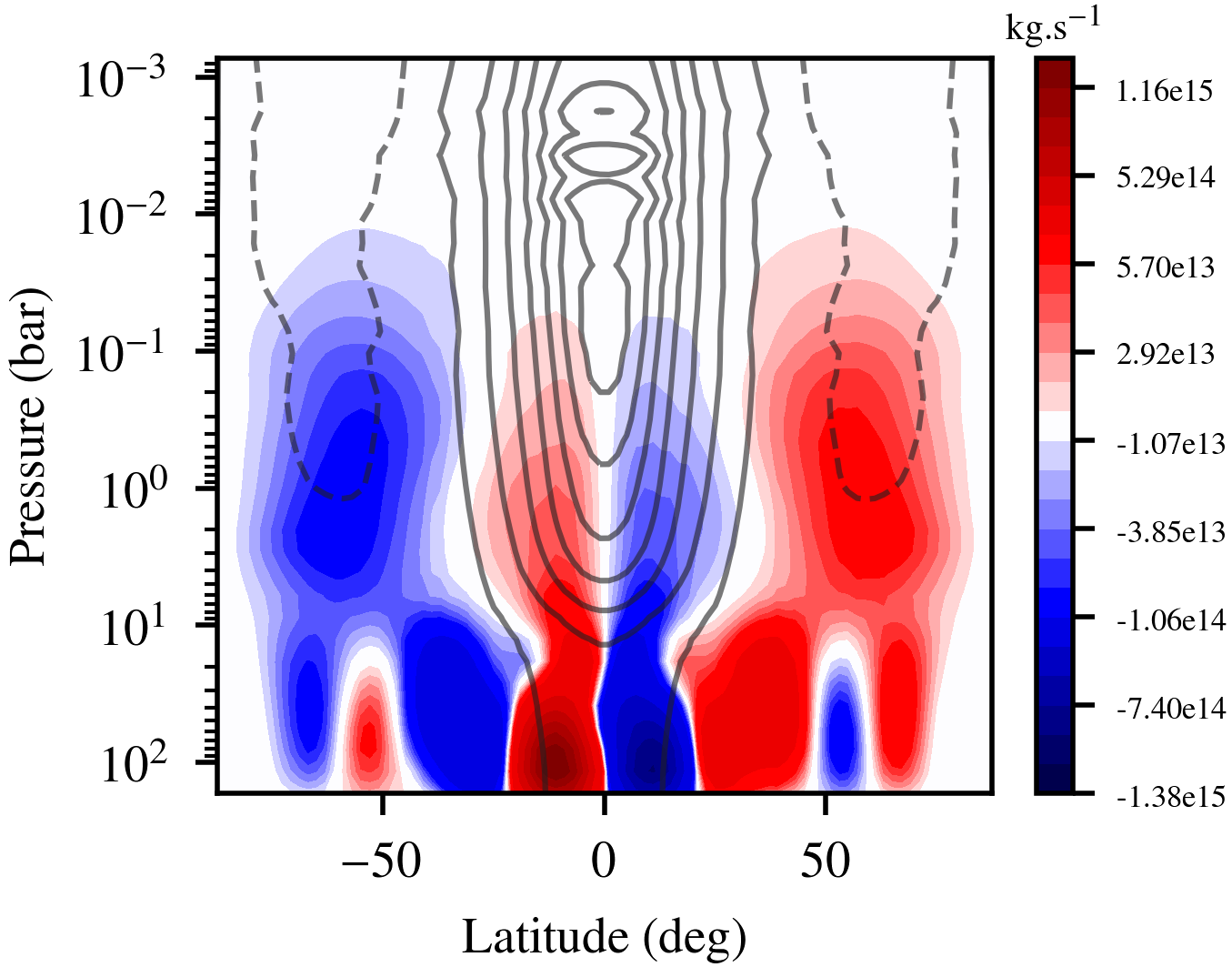}
\caption[]{$\log(\tau_{220})=20$  \label{fig:deep_tau_forcing_streams_B} }
\end{centering}
\end{subfigure}
\caption[Plots showing the Streamfunction (i.e. vertical mass flow and Meridional Circulation) for models with either weak or strong cooling below 10 bar.]{{\ed Zonally and temporally-averaged (over a period of $\approx30$ years) stream-function for for two models with either a relatively `strong' ({\it G} - left - $\log(\tau_{220})=15$) or very weak ({\it H} - right - $\log(\tau_{220})=20$) isothermal relaxation (cooling) in the deep atmosphere $\left(P>\textrm{10 bar}\right)$. Clockwise circulations on the meridional plane are shown in red and anticlockwise circulations are shown in blue. Additionally the zonally and temporally averaged zonal wind is plotted in black (solid = eastward, dashed = westward).}  \label{fig:deep_tau_forcing_streams} }
\end{centering}
\end{figure*}  

It is unlikely that the atmosphere will suddenly turn thermally inert at pressures greater than $10$ bar. Rather, we expect the thermal time-scale will gradually increase with increasing pressure. In this section, we examine the sensitivity of the deep atmospheric flows, circulations, and thermal structure to varying levels of Newtonian cooling. Additionally we are motivated to quantify the maximum amount of Newtonian cooling under which the deep atmosphere is still able to maintain a deep adiabat.\\
To explore this, we consider five models each with different cooling time-scales at the bottom of the atmosphere {\ed (i.e. five different values of $\log(\tau_{220})$)}. From this, we can then linearly interpolate the relaxation time-scale in $\log(P)$ space between $10$ and $220$ bar. The resultant profiles are plotted in \autoref{fig:deep_tau_stability_profiles},  and can be split into three distinct groups: 1) The relaxation profile with $\log(\tau_{\textrm{220}})=7.5$ (model {\it E}) represents a case with rapid Newtonian cooling that does not decrease with increasing pressure; 2) The case $\log(\tau_{\textrm{220}})=11$ (model {\it F}) is a simple linear continuation of the relaxation profile we use between $P=\textrm{1 bar and 10 bar}$. It is the simplest possible extrapolation of the upper atmosphere thermal time-scale profile, and likely represents the strongest realistic forcing in the deep atmosphere; 3) The remaining relaxation profiles,  $\log(\tau_{\textrm{220}})=\textrm{15, 20, 22.5}$ (models {\it G, H} and {\it I}), represent heating and/or cooling processes that get progressively slower in the deep atmosphere, in accordance with expectations born out from 1D atmospheric models of hot Jupiter atmospheres (see, for example, \citealt{2005A&A...436..719I}).\\
The results we obtained are summarised by the $T$--$P$ profiles we plot in \autoref{fig:deep_tau_stability}. For low levels of heating and cooling in the deep atmosphere (models {\it G, H} and {\it I}), the results are almost indistinguishable from models {\it A} and {\it B}, with only a decrease in the outer atmosphere connection temperature of a few Kelvin in model {\it G}. We find a more significant reduction in the temperature of the $T$--$P$ when we investigate model {\it F}, in which we set $\log(\tau_{\textrm{220}})=11$. In particular, there is a deepening of the connection point between the outer atmosphere and the deep adiabat, which only becomes apparent for $P>20$ bar in this model. This result suggest that model {\it F} falls near the pivot point between models in which the deep atmosphere is adiabatic and those that relax toward the imposed temperature profile. This is confirmed by model {\it E}, in which $\tau_{\textrm{220}}=7.5$, where we find that the deep adiabat has been rapidly destroyed (in $<\textrm{30 years}$), such that the deep $T$--$P$ profile corresponds to the imposed cooling profile throughout the atmosphere. This occurs because the Newtonian time-scale has become smaller than the advective time-scale, which means that the imposed temperature profiles dominates over any dynamical effects. \\
Before closing this section, let us briefly comment on the meridional
circulation profiles obtained in those models that converge onto a similar
deep adiabatic temperature profile (models {\it G, H} and {\it I}). For all of
them, we recover the same qualitative structure we found for model {\it A},
characterised by meridional cells of alternating direction that extend from the
deep atmosphere to the outer regions. The finer details of the
circulations, however, differ from the ones seen in model {\it A}. This is illustrated in
\autoref{fig:deep_tau_forcing_streams} which displays the meridional circulation
and zonal flow profiles for models {\it G}
(\autoref{fig:deep_tau_forcing_streams_A}) and {\it H}
(\autoref{fig:deep_tau_forcing_streams_B}). As the Newtonian cooling becomes
faster in the deep atmosphere, the number of meridional cells increases (see
also \autoref{fig:vertical_streamfuction_mass_flow}), to the point that, in
model {\it E}, no deep meridional circulation cells exists and the deep
circulation profile is essentially unstructured.
Despite these differences in the shape of the meridional circulation,
the steady state profiles obtained in these simulations in the deep atmosphere is again an adiabatic PT profile provided the Newtonian cooling is not (unphysically)
strong.

%Note that preliminary studies of the formation of a deep adiabat from an isothermal initial state suggests that models with deep cooling take longer to reach a steady-state. The exact cause of this is unknown, but it can likely be linked to a combination of energy losses from deep cooling, and a reduction in vertical potential temperature transport as the circulation profile weakens. Future work should examine the robustness and origin of these changes and the sensitivity to the numerical scheme being used. %We also find that, as the strength of the deep cooling is increased, the strength of the vertical transport of potential temperature has been somewhat suppressed. This can be seen by looking at the energy evolution of the deep atmosphere. Future work should examine the robustness and origin of these changes and the sensitivity to the numerical scheme being used.

\section{Conclusion and discussion} 
\label{sec:conclusion}

\begin{figure}[ht] %
\begin{centering}
\includegraphics[width=0.9\columnwidth]{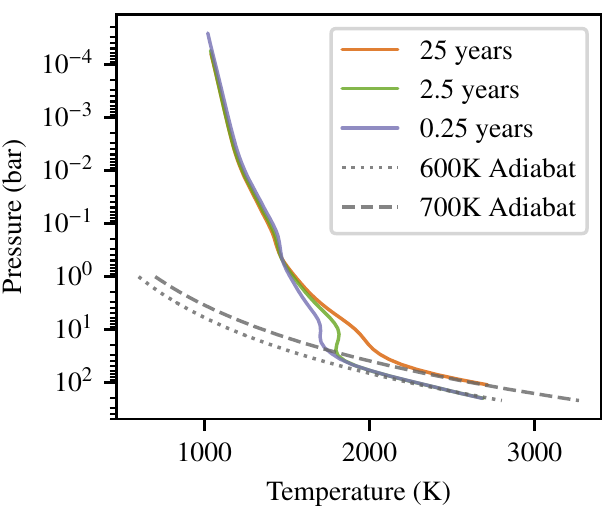} 
\caption[Stability of the deep adiabat against forcing to an
  Isothermal state.]{ Evolution of the sub-stellar point
  (i.e. day-side) Temperature-Pressure profile in a simulation
  (detailed in \citealt{2016A&A...595A..36A}) calculated using the Met
  Office GCM, the Unified Model, \citep{2014GMD.....7.3059M} and
  including a robust two-stream radiation scheme
  \citep{2014A&A...564A..59A}. Here we show snapshots of the T-P
  profile at 0.25 (purple), 2.5 (green), and 25 (orange) Earth years,
  along with two example adiabats (grey dotted and dashed lines)
  designed to show how the deep atmosphere gets warmer and connects to
  steadily warmer adiabats as the simulation progresses. Note that
  this progression is, at the end of the simulated time, ongoing
  towards a deep, hot, adiabat, albeit at an increasingly slow
  rate.   \label{fig:Metoffice_GCM_fig} }
\end{centering}
\end{figure}

\subsection{Conclusions of the simulation results}

By carrying out a series of 3D GCM simulations of irradiated atmospheres, we have shown in the present paper that:
\begin{itemize}
\item If the deep atmosphere is initialised on an adiabatic PT profile, it
  remains, as a steady state, on this profile,
\item If the deep atmosphere is initialised on a too hot state, it rapidly
  cools down to the same steady state adiabatic profile,
\item If the deep atmosphere is initialised on a too cold state, it slowly
  evolves towards the steady state adiabatic profile.
\end{itemize}
Furthermore, in all the above cases, the deep adiabat forms at lower pressures that those at which we would expect, from 1D models, the atmosphere to be convectively unstable. 
We have also shown that this steady-state adiabatic profile is stable to changes
in the deep Newtonian cooling and is independent of the details of the flow
structures, provided that the velocities are not completely negligible.
The hot adiabatic deep atmosphere is the natural final outcome of the simulations, for various
resolutions, even though the time-scale to reach steady-state is
longer at higher resolution when starting from a too cold initial state.

When the simulations are initialised on a too cold profile, the time-scale to reach the
steady state is of the order of $t\sim\textrm{1000 years}$, explaining why the
formation of a deep adiabat has not been seen in previous GCM studies: this time-scale
far outstrips the time taken for the outer
atmosphere to reach an equilibrium state ($t<\sim\!\textrm{$1$ year}$ for
$P<\textrm{1 bar}$). As a result, the vast majority of published GCM models
only contain a {\it partially evolved deep atmosphere}, the structure of which is
directly comparable to the early outputs of our isothermally initialised
calculation. Examples of this early evolution of the deep atmosphere towards a
deep adiabat (as seen in the early outputs plotted in
\autoref{fig:time_evo_from_iso}) include Figure 6 of \citet{2010ApJ...714.1334R}
(where the deep temperature profile shows signs of heating from its initial
isothermal state, albeit only on the irradiated side of the planet), Figure 7 of
\citet{2016A&A...595A..36A} (where we see a clear shift from their initially
isothermal deep atmosphere towards a deep adiabat), and Figure 8 of
\citet{2015ApJ...801...86K} (where we again see a temperature inversion and a
push towards a deep adiabat for Wasp-43b).  It is tempting to think that if
these simulations were run longer, they would evolve to a similar, deep
adiabatic structure (with a corresponding increase in the exoplanetary
radius). In order to investigate this possibility, we have extended the model of
\citealt{2016A&A...595A..36A}, run with the Unified Model of the Met Office (which includes a robust two-stream radiation scheme that replaces the Newtonian Cooling in our models), for
a total of $\approx 25$ Earth years. The results 
are shown in \autoref{fig:Metoffice_GCM_fig}, where we plot the
pressure-temperature profile at three different times, along with examples of
the approximate deep adiabat that best matches each snapshot. We see
that the deep atmosphere rapidly converges towards a deep adiabat with
further vertical advection of potential temperature warming up this adiabat as the
simulation 
goes on. Since this process keeps going on during the simulation, the
result not only reinforces our conclusions but suggests that our primary
Newtonian cooling profile represents a reasonable approximation of the incident
irradiation and radiative loss.\\
The results obtained in the present simulations suggest that future hot Jupiter atmosphere studies should be
initialised with a hot, deep, adiabat starting at the bottom of the surface
irradiation zone ($P\sim\!\!\textrm{10 bar}$ for HD209458b). Furthermore, in a
situation where the equilibrium profile in the deep atmosphere is uncertain,
we suggest that this profile should be initialised with a hotter adiabat
than expected rather than a cooler one. The simulation should then be run long enough for the deep atmosphere to reach equilibrium. This is in
agreement with the results of \citet{2016A&A...595A..36A}, who also suggested that future GCM models
should be initialised with hotter profiles than currently considered. 
For instance, recent 3D simulations of HD209458b have been
initialised with a hotter interior T-P profile (for example, one of their models is initialised with an isotherm that is $800$~K hotter than typically used in GCM studies, thus bringing the deep atmosphere closer towards its deep adiabat equilibrium temperature), and show
important differences, on the time-scales considered, between the internal dynamics obtained with this set-up, and the ones obtained with a 
cooler, more standard, deep atmospheric profile (see,
\citealt{2018MNRAS.481..194L,2018A&A...615A..97L,  2019MNRAS.tmp.1746L}). 
Using aforementioned more correct atmosphere initial profiles should not only bring these models towards
a more physical hot Jupiter parameter regime (with then a correct inflated radius),
but also provide a wealth of information on how the deep adiabat responds to
changes in parameter and computational regime.

\subsection{Evolution of highly irradiated gas giants}

The results obtained in the present GCM simulations have strong implications
for our understanding of the evolution of highly irradiated gas giants. As just mentioned, we first
emphasise that simulations initialised from a too cold state are not relevant for the evolution of inflated hot Jupiters (although it could be of some interest for re-inflation, but this is beyond the scope of this paper). Indeed, inflated hot Jupiters are
primarily in a hot initial state and, as far as the evolution is concerned, only the steady state of
the atmosphere matters. The shorter timescales needed
to reach this steady state are irrelevant for the evolution (with a typical Kelvin Helmholtz timescale of $\sim 1\textrm{Myr}$).\\
As shown in the present simulations, provided they are run long enough, hot Jupiter
atmospheres converge at depth, i.e. in the optically thick domain, to a hot adiabatic steady-state profile without the need to invoke any dissipation
mechanism such as ohmic, or kinetic energy, dissipation. These 3D dynamical calculations thus confirm the 2D
steady-state calculations of \citetalias{2017ApJ...841...30T}. Importantly enough,
the transition to an adiabatic atmospheric profile occurs at lower pressures than the ones at which the medium is
expected to become convectively unstable (thus adiabatic according to the Schwarzchild criterion). This means that the
planet lies on a hotter internal entropy profile than suggested by
1D irradiation models, yielding a larger radius. The mechanism of potential temperature advection in the atmosphere of irradiated planets thus provides a robust solution to the radius
inflation problem.\\
As mentioned previously, {\ed almost} all scenarios suggested so far to resolve the anomalously inflated planet problem rely on the (uncomfortable) necessity to introduce finely tuned parameters.
This is true, in particular, for all the different dissipation mechanisms, whether they involve kinetic energy, or ohmic and tidal dissipation. This is in stark contrast with the present mechanism, in which {\it entropy}
(potential temperature) is advected from the top to the bottom of the atmosphere. High entropy fluid parcels are moved from the upper to the deep atmosphere and toward high latitude while low entropy fluid parcels come from the deep atmosphere and are deposited in the upper atmosphere. This gradually changes the entropy profile until a steady state situation is obtained. Although an {\ed enthalpy} (and mass and momentum) flux is associated with this process, down to the bottom of the
atmosphere (characterised by some specific heat reservoir), this does not require a dissipative process {\ed (from kinetic, magnetic or radiative energy reservoirs into the internal energy reservoir)}. \\
{\ed In order to characterise this deep heating flux, and confirm that our hot, deep, adiabat would not be unstable due to high temperature radiative losses, we also explored the vertical enthalpy flux in our model and compared it to the radiative flux, as calculated for a deep adiabat using ATMO (\citetalias{2017ApJ...841...30T}). This analysis reveals that the vertical enthalpy flux dominates the radiative flux at all $P>1$ bar: For example, averaging over a pressure surface at $P=10$ bar, we find a net vertical enthalpy flux ($\rho c_{p}Tu_{z}$) of $ -1.04\times10^8\mathrm{erg s^{-1} cm^{-2}}$ compared to a outgoing radiative flux of $ 7.68\times10^6\mathrm{erg s^{-1} cm^{-2}}$, suggesting that any deep radiative losses are well compensated by energy (enthalpy) transport from the highly irradiated outer atmosphere. This result is reinforced by UM calculation we show in \autoref{fig:Metoffice_GCM_fig}, which intrinsically includes this deep radiative loss and show no evidence of cooling due to deep radiative effects.  }
\\
%The lack of requirement of dissipative process in the present mechanism can be understood as follows. 
This ({\ed lack of a requirement for additional dissipative processes}) is of prime importance when trying to understand the evolution of irradiated planets. Whereas dissipative processes imply an extra energy source in the evolution ($\int_M {\dot \epsilon} dm$, where ${\dot \epsilon}$ is the energy dissipation rate, to be finely tuned), to slow down the planet's contraction, there is no need for such a term in the present process. Indeed, as an {\it isolated} substellar object (i.e. without nuclear energy source) cools down, its gravitational potential energy is converted into radiation at the surface, with a flux $\sigma T_\mathrm{eff}^4$. Let us now suppose that the same
object is immersed into an isotropic medium characterised by a pressure $P=$220 bars and a
temperature $T\sim 4000$~K, typical conditions in the deep atmosphere of 51Peg-B like hot Jupiters.
{\ed Once the object's original inner adiabat (after its birth)
has cooled down to 4000 K at 220 bars, the thermal gradient between the external and internal media will be null, which essentially reduces the local convective flux and the local optically thick radiative flux to zero. Thus the cooling flux will be reduced to almost zero. At this point, the core}
cannot significantly cool any more and is simply in thermal equilibrium with the surrounding medium. Both the contraction and the cooling flux are essentially insignificant: $dR/dt\approx 0, \sigma T_\mathrm{int}^4
\approx 0$, {\ed in which we define $\sigma T_\mathrm{int}^4$ as the radiative and convective cooling flux at the interior-atmosphere boundary}. {\ed Indeed, convection will become inefficient in transporting energy and any remaining radiative loss in the optically thick deep core will be compensated by downward energy transport from the hot outer atmosphere.  }  For a highly irradiated gas giant, the irradiation flux is not
isotropic, but the combination of irradiation and atmospheric circulation will
lead to a similar situation, with a deep atmosphere adiabatic profile of $\sim$4000~K at 220 bars
for all latitudes and longitudes. Therefore, the planet's interior does not significantly cool
any more and we also have $\sigma T_\mathrm{int}^4 \approx 0$.  The evolution of the planet is stopped ($dS/dt\approx0$), or let say its cooling time is now prohibitively long, and the planet lies on a constant adiabat determined by the equilibrium between the inner and atmospheric ones at the
interior-atmosphere boundary. The situation will last as long as the planet-star characteristics will remain the same, illustrating the robustness of the potential temperature advection mechanism to explain the anomalous inflation of these bodies.\\
Therefore, the irradiation induced advection of potential temperature appears
to be the most natural and robust processes to resolve the radius-inflation puzzle. Note that, this does not exclude other processes (e.g. dissipative ones) from operating within hot Jupiter
atmospheres, but they are unlikely to be the dominant mechanisms responsible for the radius inflation.

\begin{acknowledgements}
FSM and PT would like to acknowledge and thank the ERC for funding this work under the Horizon 2020 program project ATMO (ID: 757858). NJM is part funded by a Leverhulme Trust Research Project Grant and partly supported by a Science and Technology Facilities Council Consolidated Grant (ST/R000395/1). JL acknowledges funding from the European Research Council (ERC) under the European Union's Horizon 2020 research and innovation program (grant agreement No. 679030/WHIPLASH). GC was supported by the Programme National de Planétologie (PNP) of CNRS-INSU cofunded by CNES.  IB thanks the European Research Council (ERC) for funding under the H2020 research and innovation programme (grant agreement 787361 COBOM). FD thanks the European Research Council (ERC) for funding under the H2020 research and innovation programme (grant agreement 740651 NewWorlds). BD acknowledges support from a Science and Technology Facilities Council Consolidated Grant (ST/R000395/1). \\
The authors also wish to thank Idris, CNRS, and Mdls for access to the supercomputer Poincare, without which the long time-scale calculations featured in this work would not have been possible. The calculations for Appendix A were performed using Met Office software. Additionally, the calculations in Appendix A used the DiRAC Complexity system, operated by the University of Leicester IT Services, which forms part of the STFC DiRAC HPC Facility (www.dirac.ac.uk). This equipment is funded by BIS National E-Infrastructure capital grant ST/K000373/1 and  STFC DiRAC Operations grant ST/K0003259/1. DiRAC is part of the National E-Infrastructure. \\
Finally the authors would like to thank the Adam Showman for their careful and insightful review of the original manuscript. 
%Finally F. Sainsbury-Martinez would like to thank Sam Morrell and Tim Naylor for an interesting, and enlightening, discussion of the evolution of post main-sequence stars. 
\end{acknowledgements}

\bibliographystyle{aa} 
%\interlinepenalty=10000
\bibliography{main}

\end{document}